\documentclass[nofootinbib,superscriptaddress,aps,prd,showkeys,noshowpacs,onecolumn,10pt]{revtex4}
\usepackage{eurosym}
\usepackage{graphics}
\usepackage{graphicx}
\usepackage{epsf}
\usepackage{bm}
\usepackage{amsmath,amssymb,amsfonts}
\usepackage{latexsym}
\usepackage{enumerate}

\def\be{\begin{equation}}
\def\ee{\end{equation}}

\makeatother

\begin{document}

\title{Cosmo-PINN: A Physics-Informed Neural Network for Cosmological
Reconstruction}
\author{Andronikos Paliathanasis}
\email{anpaliat@phys.uoa.gr}
\affiliation{Institute of Systems Science, Durban University of Technology, Durban 4000,
South Africa}
\affiliation{Centro de Investigaci\'on, Innovaci\'on y Creaci\'on (CIIC) \& Departamento
de Ciencias Matem\'{a}ticas y F\'{\i}sicas, Facultad de Ingenier\'ia,
Universidad Catolica de Temuco, 4813302, Temuco, Chile}
\affiliation{National Institute for Theoretical and Computational Sciences (NITheCS),
South Africa.}

\begin{abstract}
We introduce Cosmo-PINN, a Physics-Informed Neural Network for
reconstruction of the cosmological theory. In this work we demonstrate the
application of the Cosmo-PINN in the reconstruction of the dark energy
equation of state parameter $w_{DE}\left( z\right) $ directly from late-time
cosmological observations. This framework overcomes the main limitation
shared by Gaussian Process and Artificial Neural Network reconstruction
approaches, where the recovered solution is driven by the data and it is not
necessarily true that it is physically consistent, by embedding the
cosmological constraints directly into the loss function as constraints,
ensuring that the reconstructed quantities satisfy the physical laws at
every point during the training. For the training of the network, we
employed background data, and specifically the Baryon Acoustic Oscillation
from DESI DR2, the Cosmic Chronometers and three different Supernova
compilations, while we simultaneously introduce the cosmological parameters $%
H_{0},~\Omega _{m0}$ and $r_{\mathrm{drag}}$ as trained parameters. The
reconstruction shows that the trained $w_{DE}\left( z\right) $ crosses the
phantom divide around $z=0.5$ in agreement with the value obtained by the
Chevallier-Polarski-Linder model. In the quintessence scenario, the dark
energy equation of state parameter is forced to the $\Lambda $CDM limit in
the past, that is $w_{DE}\left( z\right) \rightarrow -1 $. Finally, we
demonstrate the significance of imposing the physical constraints within the
loss function by comparing the Cosmo-PINN reconstruction against a purely
data-driven neural network with the same architecture.
\end{abstract}

\date{\today}
\maketitle

\section{Introduction}

The mechanism responsible for the late-time acceleration phase of the
observable universe \cite{rr1,Teg,Kowal,Komatsu,suzuki11} remains one of the
central open problems in modern cosmology. The gravitational field describes
attractive interactions, however in order to describe the late-time
acceleration in the framework of General Relativity, it is required to
introduce a matter source with repulsive force contribution in order to
drive the cosmic acceleration. This component, which is detected only
through its gravitational effects, is referred as dark energy \cite{jo,jo1}.
The cosmological constant is the simplest theoretical motivated dark energy
candidate. The cosmological constant introduces the minimum degrees of
freedom in the gravitational field equations and until recently it was the
standard model for the description of the cosmic acceleration. The
corresponding cosmological model, the $\Lambda $CDM, is nowadays challenged
on the observational side by the cosmological tensions \cite{ten2}.
Furthermore, the cosmological constant suffers from two other major
problems, the fine tuning and the coincidence problems \cite%
{Weinberg89,Pad03,Peri08}.

A plethora of alternative proposals to the cosmological constant have been
introduced in the literature in order to address the above issues, we refer
the reader to \cite{ter3} for a systematic review. The proposed solutions
can be classified into models which modify the matter sector, such as
Chaplygin gas models \cite{cg1}, interacting dark sector \cite%
{int1,int2,int3}, scalar fields~\cite{ratra,peebles,pha,mp1,mp2,mp3,luongo}, 
$k$-essence \cite{kes}, and modifications of the gravitational action
integral, including scalar-tensor models \cite{st}, Horndeski theory \cite%
{hr}, $f\left( R\right) $-gravity \cite{fr}, teleparallel $f\left( T\right) $%
-gravity \cite{ft}, symmetric teleparallel $f\left( Q\right) $-gravity \cite%
{fq} and many others \cite{md1,md2,md3,md5,md6,md7,md8,md9,md10}.

Phenomenological approaches form an alternative and simpler framework for
the description of the cosmological data, which allows us to extract
important information about the nature of dark energy, since they allow us
to explore how dark energy is distinct from the cosmological constant
scenario \cite{cpl2,cpl4}. The most widely used phenomenological dark energy
model is the Chevallier-Polarski-Linder (CPL) model \cite{cpl1,cpl1a}. CPL
belongs to the class of $w_{0}w_{a}$CDM models, in which the dark energy
equation of state parameter is considered to be a function of the redshift
with two free variables, where $w_{0}$ usually refers to the value of the
dark energy equation of state parameter at the present, for more details we
refer the reader to \cite{par1,par4,par5,exp1A,osc1,osc3,osc5} and
references therein. Nevertheless, phenomenological parametrizations are
model-dependent because the functional form of the $w_{DE}\left( z\right) $
is imposed a priori, and the resulting observable predictions therefore
carry the imprint of the assumed function rather than of the underlying
physics.

A reconstruction approach is essential in order to extract information for
the properties of the physical theory, directly from the observational data,
without restricting the free functional forms of the theory beforehand. In
the standard Bayesian method, the parameter estimation is usually performed
under a specific assumption for the cosmological model, that is a given
functional form of the dark energy equation of state $w_{DE}\left( z\right) $%
, or the Hubble function~\thinspace $H\left( z\right) \,$, such as the $%
\Lambda $CDM model or the CPL parametrization discussed above. The
limitation of the Bayesian approach is that the conclusions made are valid
only for the specific model assumed. Various phenomenological models for the
dark energy equation of state can exhibit similar behaviour at a given
cosmological epoch and fit the observational data in a similar way, such
that the models are statistically indistinguishable \cite{anlcdm}. Moreover,
in theoretically motivated models, such as the quintessence scalar field,
there exists a plethora of scalar field potentials that share a common
attractor solution, causing all such models to predict a similar dynamical
evolution \cite{map1} which as a result leads to a similar fitting to the
background data. Two main approaches to the reconstruction problem which can
be found in the literature are the Gaussian Process (GP) \cite%
{gp1,gp2,gp3,gp4,gp5,gp6,gp7,gp8,gp9,gp10,gp11,gp12,gp13,gp14,gp15,gp16,gp17a}
and the Artificial Neural Networks (ANNs) \cite%
{an1,an2,an3,an4,an5a,an5,an6,an7,an8,an9}. GP is a model-independent
approach, however, it suffers from two main issues. The selection of the
kernel may affect the reconstructed physical parameters \cite{gp17,gp18},
while the GP suffers from overfitting issues. On the other hand, the ANN
approach provides a fully data-driven, model-independent reconstruction
framework; however, there is no guarantee that the trained solution
satisfies the field equations governing the cosmological dynamics, since the
physical laws are not incorporated into the training. While some other model
independent approaches for the reconstruction are discussed in \cite%
{mpp1,mpp2,mpp3,mpp4,mpp5}.

Physics-Informed Neural Networks (PINNs) \cite{pinnp}, in which the physical
laws, that is, the field equations in cosmological studies, are embedded
directly into the loss function offer a natural approach to overcome this
limitation during the reconstruction \cite{pinnbook}. PINNs have been
applied previously in cosmological studies. In \cite{pnn1} a PINN was used
to solve the background cosmological field equations both in General
Relativity and in modified theories of gravity. Furthermore, in \cite%
{pnn2,pnn3}, the PINN approach was employed to reconstruct the Hubble
function within the Tsallis and Barrow holographic dark energy models, where
the field equations were embedded directly into the network training.
Moreover, in \cite{pnn4}, a PINN framework was introduced to perform
parameter estimation for parametric dark energy models. On the other hand,
in \cite{pnn5} a PINN network was applied to solve the linear matter
perturbations. In all these approaches the PINN applied within a specific
cosmological model, and the reconstructed solution is therefore
model-dependent. On the other hand, a PINN was introduced recently in \cite%
{pnn6} for the reconstruction of the supernova absolute magnitude.

In this work we introduce Cosmo-PINN, a framework that combines the ANN
reconstruction with the physics enforcement of PINNs. Cosmo-PINN imposes the
cosmological field equations as constraints in the loss function, ensuring
that the reconstructed observables are consistent with the observational
data at each point and with the underlying physical law throughout the
training domain. The framework can be applied to any cosmological model. The
advantage of considering the field equations in the training becomes
essential when the field equations do not admit a closed-form solution,
which is the case for the majority of cosmological scenarios, such as
interacting models, non-minimally coupled scalar-tensor theories and
modified theories of gravity which discussed above. In such cases the
parameter estimation requires the numerical integration of the field
equations at every evaluation of the likelihood. PINNs overcome this
limitation by providing the solution of the dynamical equations and
performing the inference within a single training process. Specifically, we
focus on the reconstruction of the dark energy equation of state parameter $%
w_{DE}\left( z\right) $\ and of the related cosmological parameters. The
structure of the paper is as follows.

In Section \ref{sec2} we present the cosmological framework under
consideration, namely the spatially flat Friedmann-Lema\^{\i}%
tre-Robertson-Walker universe consisting of radiation, baryons, dark matter
and a dark energy fluid, characterized by a parametric equation of state
parameter $w_{DE}\left( z\right) $. The basic architecture features of the
Cosmo-PINN framework are introduced in Section \ref{sec3}. In Section \ref%
{sec4} we demonstrate the application of the Cosmo-PINN for the
reconstruction of the dark energy equation of state parameter from the
cosmological data by imposing the cosmological field equations as
constraints in the loss function. We examine two scenarios, the $%
w_{DE}\left( z\right) $ to be unbound, and the quintessence scenario in
which $w_{DE}\left( z\right) \geq -1$. Furthermore, in Section \ref{sec5} we
perform the same reconstruction by considering a Neural Network with the
same architecture but without imposing the physical constraints. In Appendix %
\ref{stabtest} we present two tests for the stability of the Cosmo-PINN,
while in appendix \ref{append}, we extend the Cosmo-PINN framework to the
reconstruction of cosmological parameters in teleparallel$f\left( T\right) $%
-gravity. Finally, in Section \ref{sec6} we summarize our results and
discuss future applications of the Cosmo-PINN framework.

\section{FLRW Cosmology and Dark Energy}

\label{sec2}

On very large scales, the universe is observed to be both isotropic and
homogeneous. The geometry which defines the distances in cosmological scales
is described by the spatially flat FLRW metric, with line element%
\begin{equation}
ds^{2}=-dt^{2}+a^{2}\left( t\right) \left( dx^{2}+dy^{2}+dz^{2}\right) ,
\label{cc.01}
\end{equation}%
Function $a\left( t\right) $ is the scale factor which describes the radius
of the three-dimensional hypersurface, and $H=\frac{\dot{a}}{a}$ is the
Hubble function which describes the expansion history, and an overdot means
differentiation with respect to the time parameter $t$, i.e. $\dot{a}=\frac{%
da}{dt}$.

Within the framework of General Relativity, the dynamical evolution of scale
factor $a\left( t\right) $ is given by the Einstein field equations\footnote{%
We consider units where $8\pi G=1$ and $c=1$.}%
\begin{align}
3H^{2}& =\rho ,  \label{c.02} \\
-2\dot{H}-3H^{2}& =p  \label{c.03}
\end{align}%
where $\rho $ and~$p$ are the energy density and pressure components for the
cosmic fluid, with energy momentum tensor%
\begin{equation}
T_{\mu \nu }=\left( \rho +p\right) u_{\mu }u_{\nu }+pg_{\mu \nu },
\label{c.04}
\end{equation}%
and $u^{\mu }=\delta _{t}^{\mu }\,$\ with $u^{\mu }u_{\mu }=-1$,~be the
comoving observer. The equation of state parameter for the cosmological
fluid is $w_{tot}=\frac{p}{\rho }$, and the deceleration parameter is given
by the expression $q=\frac{1}{2}\left( 1+3w_{tot}\right) $.

Furthermore, by differentiating equation\ (\ref{c.02}) and by using equation
(\ref{c.03}), we obtain the equation of motion for the cosmic fluid 
\begin{equation}
\dot{\rho}+3H\left( \rho +p\right) =0.  \label{c.05}
\end{equation}%
The latter equation is the conservation law of energy for the cosmic fluid
and it can be derived from the application of the Bianchi identity for the
Einstein field equations, that is, in tensor form equation (\ref{c.05})
reads $\nabla _{\nu }T_{~}^{\mu \nu }=0$, in which$~\nabla _{\mu }$ defines
covariant derivative with respect to the Levi-Civita connection for the FLRW
geometry (\ref{cc.01}).

The cosmological fluid is assumed to be composed of the following
components, the radiation $\rho _{r}$,~$p_{r}=\frac{1}{3}\rho _{r}$, the
pressureless baryonic matter $\rho _{b},~p_{b}=0$, the cold dark matter $%
\rho _{m},~p_{m}=0$ and the dark energy component which drives the cosmic
acceleration, with energy density $\rho _{DE}$ and equation of state
parameter $w_{DE}$, that is, $p_{DE}=w_{DE}\rho _{DE}$. \ Thus, the energy
momentum tensor (\ref{c.04}) is expressed as%
\begin{equation}
T_{\mu \nu }=\left( \rho _{m}+\rho _{b}+\frac{4}{3}\rho _{r}+\left(
1+w_{DE}\right) \rho _{DE}\right) u_{\mu }u_{\nu }+\left( \frac{1}{3}\rho
_{r}+w_{DE}\rho _{DE}\right) g_{\mu \nu }.  \label{c.06}
\end{equation}%
Therefore, the equation of state parameter for the cosmic fluid is expressed
as $w_{tot}=\frac{\frac{1}{3}\rho _{r}+w_{DE}\rho _{DE}}{\rho },$ or
equivalently,%
\begin{equation}
w_{tot}=\frac{1}{3}\Omega _{r}+w_{DE}\Omega _{DE},  \label{c.07}
\end{equation}%
in which the components $\Omega _{i}=\frac{\rho _{i}}{3H^{2}}$ describe the
energy density for the fluid $\rho _{i}$, and from equation\ (\ref{c.02})
the following algebraic constraint expression follows%
\begin{equation}
\Omega _{m}+\Omega _{b}+\Omega _{r}+\Omega _{DE}=1.  \label{c.08}
\end{equation}

With the use of the latter variables, equation (\ref{c.03}) reads%
\begin{equation}
\frac{d\ln H}{d\ln a}=-\frac{3}{2}\left( 1+\frac{1}{3}\Omega
_{r}+w_{DE}\left( z\right) \Omega _{DE}\right) ,  \label{c.09}
\end{equation}

Assuming that the components constituting the cosmic fluid interact only
gravitationally, then from the conservation law of energy (\ref{c.05}) it
follows%
\begin{eqnarray}
\dot{\rho}_{m}+3H\rho _{m} &=&0,  \label{c.10} \\
\dot{\rho}_{b}+3H\rho _{b} &=&0,  \label{c.11} \\
\dot{\rho}_{r}+4H\rho _{r} &=&0,  \label{c.12} \\
\dot{\rho}_{DE}+3H\left( 1+w_{DE}\right) \rho _{DE} &=&0.  \label{c.13}
\end{eqnarray}%
From where we derive%
\begin{equation}
\rho _{m}=\rho _{m,0}\left( \frac{a}{a_{0}}\right) ^{-3},~\rho _{b}=\rho
_{b,0}\left( \frac{a}{a_{0}}\right) ^{-3},~\rho _{r}=\rho _{r,0}\left( \frac{%
a}{a_{0}}\right) ^{-4}  \label{c.14}
\end{equation}%
and%
\begin{equation}
\rho _{d}=\rho _{d,0}\exp \left( -3\int_{a}^{a_{0}}\frac{1+w_{DE}\left(
\alpha \right) }{\alpha }d\alpha \right) .  \label{c.15}
\end{equation}%
in which $a_{0}$ is the value of the scale factor at the present, $a\left(
t_{0}\right) =a_{0}$, and parameters $\rho _{i,0}$ are integration constants
and define the energy density of the fluid at the present time. The energy
density for each fluid at the present is defined as $\Omega _{i,0}=\frac{%
\rho _{i,0}}{3H_{0}^{2}}$, $H_{0}$ is the Hubble constant.

Therefore, the Hubble function is given by the expression%
\begin{equation}
\left( \frac{H\left( a\right) }{H_{0}}\right) ^{2}=\Omega _{r0}\left( \frac{a%
}{a_{0}}\right) ^{-4}+\Omega _{b0}\left( \frac{a}{a_{0}}\right) ^{-3}+\Omega
_{m0}\left( \frac{a}{a_{0}}\right) ^{-3}+\Omega _{d0}\exp \left(
-3\int_{a}^{a_{0}}\frac{1+w_{DE}\left( \alpha \right) }{\alpha }d\alpha
\right) ,  \label{c.16}
\end{equation}%
where the energy densities $\Omega _{i,0}$ satisfy the algebraic constraint (%
\ref{c.08}).

In the case where $w_{DE}\left( a\right) =-1$, the dark energy component
becomes a constant, such that the Hubble function (\ref{c.16}) describes the
so-called $\Lambda $CDM. The nature of the equation of state parameter
affects dramatically the cosmic evolution. The equation of state parameter
for the CPL model is a linear function on the scale factor, that is, $%
w_{DE}^{CPL}\left( z\right) =w_{0}+w_{a}\left( 1-\left( \frac{a}{a_{0}}%
\right) \right) $ or, in terms of the redshift $z=\frac{1}{a}-1$, it follows 
$w_{DE}^{CPL}\left( z\right) =w_{0}+w_{a}\left( \frac{z}{1+z}\right) $,
which can cross the phantom divide line.

On the other hand, from a theoretical perspective, scalar fields along with
modified theories of gravity have been widely used to describe the dark
energy component and to connect the early inflationary epoch with the
late-time accelerated expansion. The quintessence scalar field stands as one
of the well-studied and simplest dark energy candidates \cite{ratra,peebles}%
. In this theory the equation of state parameter is bounded within the range 
$w_{DE}\in \left[ -1,1\right] $, with the lower limit being that of the
cosmological constant.

Since the present analysis is restricted to late-time observational data,
the contribution of radiation is omitted in what follows, as its effect on
the cosmic fluid at late times is negligible.

\section{Cosmo-PINN: A Physics-Informed Neural Network for Cosmological
Reconstruction}

\label{sec3}

We introduce a Physics Informed Neural Network (PINN) for the study of the
dark energy problem, which we call Cosmo-PINN.

PINNs are Neural Networks (NN) that are trained to solve supervised learning
tasks while respecting any given laws of physics described \cite{pinnp}.
Thus, within the framework of Cosmo-PINN the cosmological field equations
are embedded directly within the loss function as a constraint. Thus the
learning solution is not only a data-fitting reconstruction but is governed
by a physical law, ensuring that the reconstructed cosmological history is
consistent with the given gravitational theory at every point of the
redshift domain. Cosmo-PINN is an optimization-based reconstruction
framework, it provides a single optimal solution that minimizes the total
loss. Posterior uncertainties on the reconstructed functions are obtained by
sampling around this optimum solution.

\subsection{Architecture}

The Cosmo-PINN architecture is designed to reconstruct the Hubble function $%
H\left( z\right) $ and the free functions of the theory, such as the
equation of state parameter $w_{DE}\left( z\right) $, the $\rho _{DE}\left(
z\right) $. Moreover, the cosmological parameters $H_{0}$,~$\Omega _{m0}$, $%
r_{\mathrm{drag}}$ are simultaneously inferred within the same framework.
The input to the network is the redshift scalar $z\in \left[ 0,z_{\max }%
\right] $, in which $z_{\max }$ is the maximum value for the redshift as
provided by the datasets. Prior to entering the network, the redshift is
mapped onto the unit interval via the normalized transformation $x=\frac{z}{%
z_{\max }}$, such that $x\in \left[ 0,1\right] .$

The architecture of the Cosmo-PINN is illustrated in Fig. \ref{fig0}.

\subsubsection{Physical law}

\begin{figure}[t]
\centering\includegraphics[width=1\textwidth]{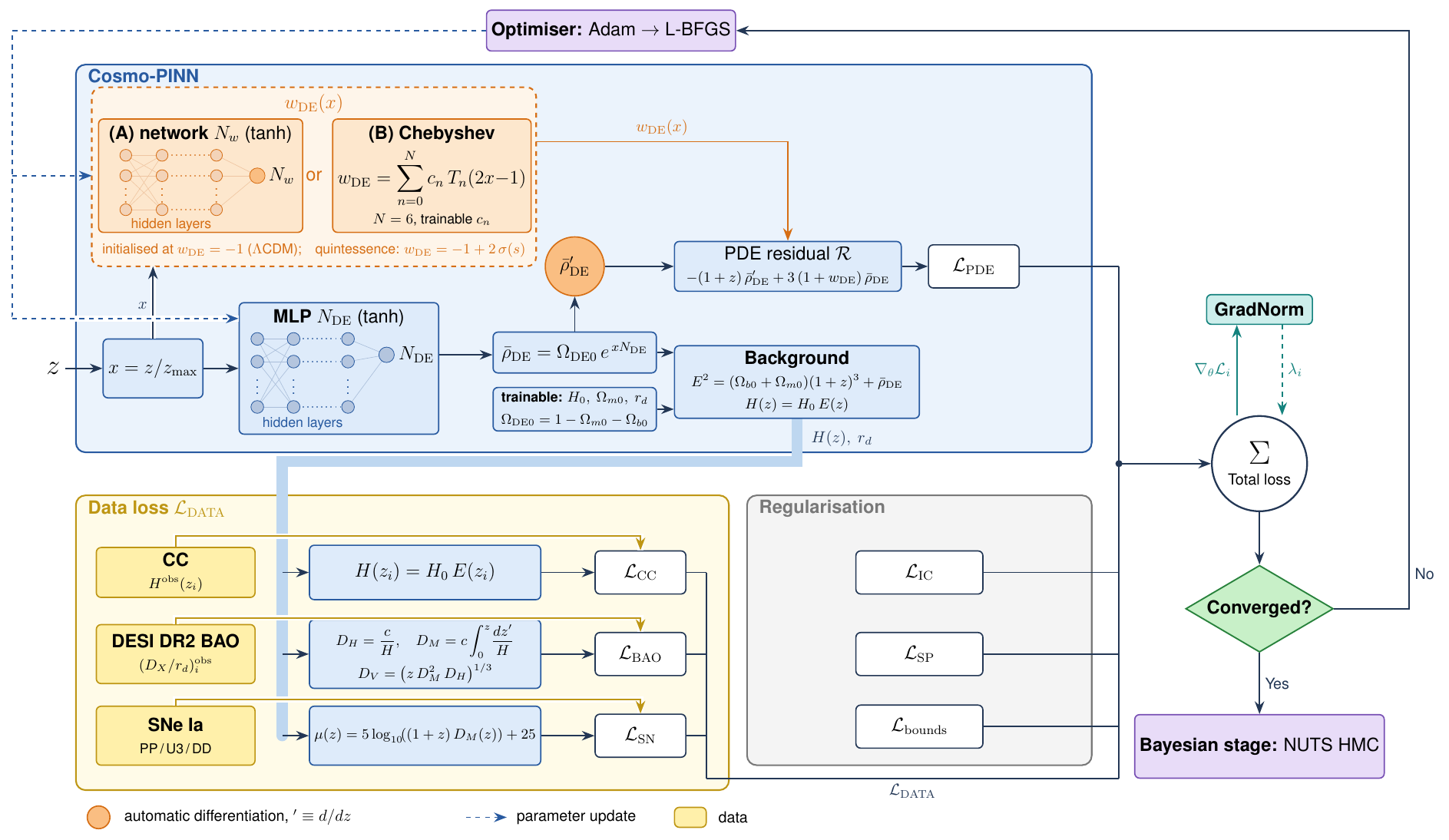}
\caption{Architecture of the Cosmo-PINN. The blue box includes the networks
that describe the physical quantities and the physical laws. The loss terms
related to the cosmological data, are presented in the yellow box, while the
gray box includes the loss terms related to the initial conditions, the
physical bounds, and the smoothness penalties. All loss terms are summarized
to the total loss function, whose weights are adapted dynamically by the
GradNorm method. The converged solution is then applied as starting point
for the Hamiltonian Monte Carlo sampling.}
\label{fig0}
\end{figure}

The full set of physical quantities is trained under the requirement of
physical consistency. In order to guarantee the trained solution possesses
the initial condition at the present time we~express the reduced dark energy
density $\bar{\rho}_{DE}\equiv \rho _{DE}/(3H_{0}^{2})$ as 
\begin{equation}
\bar{\rho}_{DE}\left( z\right) =\bar{\rho}_{DE0}\exp \left[ x\,N_{DE}(x)%
\right] ,~
\end{equation}%
where $N_{DE}(x)$ is the output of the neural network. The exponential
ansatz enforces the initial condition $\bar{\rho}_{DE0}=\Omega _{DE0}$, and
the requirement $\bar{\rho}_{DE}>0$ at every point. This is illustrated in
the blue box in Fig. \ref{fig0}.

For the equation of state parameter $w_{DE}\left( x\right) $\ we consider
two different representations. In the first approach $w_{DE}\left( x\right) $%
\ is the output of a second neural network, 
\begin{equation}
w_{DE}\left( x\right) =N_{w}\left( x\right) ,
\end{equation}%
\ while for the second approach we employ a Chebyshev polynomial expansion 
\cite{cheb1} 
\begin{equation}
w_{DE}(x)=\sum_{n=0}^{N}c_{n}^{(w)}\,T_{n}(2x-1)
\end{equation}%
with trainable coefficients $c_{n}^{(w)}$. The Chebyshev expansion provides
an analytic closed-form expression for the reconstructed $w_{DE}\left(
x\right) $. Furthermore, the Chebyshev polynomials can be used in order to
eliminate high-frequency oscillations, related to overfitting and enforce
smoothness on the reconstructed function. In addition, the Chebyshev
expansion introduces fewer degrees of freedom in the training, in comparison
to the neural network approach. As we shall see in the following, the two
approaches lead to statistically equivalent results for the late-time
reconstruction.

For the both approaches the coefficients are initialized at the $\Lambda $%
CDM point, $c_{0}^{(w)}=-1$\ and $c_{n}^{(w)}=0$\ for $n\geq 1$, so training
begins from $w_{DE}=-1$. Hence, the network is trained to explore deviations
from the $\Lambda $CDM limit. The two representations correspond to boxes
(A) and (B) within the orange box of Fig. \ref{fig0}

The matter components are pressureless and obey the exact conservation laws,
so they require no trained residual, 
\begin{equation}
\Omega _{m}(z)=\Omega _{m0}(1+z)^{3},~\Omega _{b}(z)=\Omega _{b0}(1+z)^{3},
\end{equation}%
with constraint $\Omega _{DE0}=1-\Omega _{m0}-\Omega _{b0}$. Thus the Hubble
function is derived from the first Friedmann equation 
\begin{equation}
E\left( z\right) =\left( \frac{H(z)}{H_{0}}\right) ^{2}=\Omega
_{m0}(1+z)^{3}+\Omega _{b0}(1+z)^{3}+\bar{\rho}_{DE}(z).
\end{equation}

The $\bar{\rho}_{DE}\left( x\right) $ and the $w_{DE}(x)$ are related
through the conservation equation (\ref{c.13}) 
\begin{equation}
\mathcal{R}:\frac{1+x\,z_{\max }}{z_{\max }}\,\frac{d\bar{\rho}_{DE}}{dx}%
-3\left( 1+w_{DE}(x)\right) \,\bar{\rho}_{DE}(x).  \label{rr1}
\end{equation}%
The derivatives are computed by automatic differentiation through the
network graph. The PDE loss over the $N_{c}$ collocation points is defined as

\begin{equation}
\mathcal{L}_{\mathrm{PDE}}=\frac{\lambda _{\mathrm{PDE}}}{2N_{c}}%
\sum_{i=1}^{N_{c}}\!\mathcal{R}^{2}(x_{i}).  \label{rr2}
\end{equation}%
This loss component imposes the physical law, see the PDE residual and
Background boxes in Fig. \ref{fig0}.

The Chebyshev expansion is not introduced in order to obtain a closed-form
expression for $w_{DE}\left( x\right) $, since the dark energy density is
reconstructed through the neural network and the data. The Chebyshev
polynomials have been introduced to remove the high-frequency components
which are barely visible in \thinspace $N_{DE}\left( x\right) $\ and enter
the $w_{DE}\left( x\right) $\ via the residual (\ref{rr1}).

\subsubsection{Initial conditions}

As far as the baryons are concerned, we consider the energy density $\Omega
_{b0}=\frac{\omega _{b}\,}{H_{0}^{2}}$, where $\omega _{b}$ is the value
obtained by the Planck 2018 collaboration. On the other hand, regarding the
free parameters $H_{0},~\Omega _{m0}$ and $r_{\mathrm{drag}}$ we introduce
them into the total loss via the expression%
\begin{equation}
\mathcal{L}_{\mathrm{IC}}=\frac{\lambda _{H_{0}}}{2}\left( \frac{H_{0}^{%
\mathrm{PINN}}-H_{0}^{Planck}}{\sigma _{H_{0}}}\right) ^{2}+\frac{\lambda
_{\Omega _{m0}}}{2}\left( \frac{\Omega _{m0}^{\mathrm{PINN}}-\Omega
_{m0}^{Planck}}{\sigma _{\Omega _{m0}}}\right) ^{2}+\frac{\lambda _{r_{%
\mathrm{drag}}}}{2}\left( \frac{r_{\mathrm{drag}}^{\mathrm{PINN}}-r_{\mathrm{%
drag}}^{Planck}}{\sigma _{r_{\mathrm{drag}}}}\right) ^{2}.
\end{equation}%
These terms act as soft priors that anchor the learnable cosmological
parameters. In this work we consider priors near the Planck 2018 \cite%
{planck2018} values. The reason for imposing the latter anchors is that the $%
\rho _{DE}\left( z\right) $ and the $\Omega _{m}(z)$, are not independent
within the network. The observational data constrain only the total
expansion history, thus if $\Omega _{m0}$\ is left entirely free the network
can settle into a solution in which it effectively reconstructs the equation
of state parameter of the total cosmological fluid, and not of the dark
energy component. Hence, the soft prior on $\Omega _{m0}$\ break the
degeneracy in the reconstructed solution. \ In the following we set $\lambda
_{H_{0}},~\lambda _{r_{\mathrm{drag}}}$\ to be $10^{-4}$, while the $\lambda
_{\Omega _{m0}}=10^{-2}$. These values are small, a $1\sigma $\ deviation
from $\Omega _{m0},~H_{0}$\ will contribute as $5\times 10^{-3}\,$\ and $%
5\times 10^{-5}$\ in the loss function respectively. The priors do not drive
the cosmological parameters to given values, but they prevent the network to
explore unphysical regions of solutions within the parametric space. For the 
$H_{0}$, the prior mainly speeds up convergence of the training rather than
constraining $H_{0}$, which is determined by the data, and specifically from
the CC. However, for the $\Omega _{m0}$\ parameter the prior is required,
since the $\rho _{DE}\left( z\right) $\ and the matter term $\Omega
_{m0}\left( 1+z\right) ^{3}$\ are not independent, but they are related via
the Hubble function which is constrained by the data, see gray box in Fig. %
\ref{fig0}.

\subsubsection{Smoothness penalties and physical boundaries}

We introduce a $\mathcal{L}_{\mathrm{bounds}}$ component in the loss
function, in order to impose strong boundary conditions on the trained
functions such that the obtained solution satisfies $\Omega _{m}\in \left[
0,1\right] $, or in the case of quintessence the $w_{DE}$ is within the
range $w_{DE}\in \left[ -1,1\right] $. In particular in the quintessence
limit we impose that the $w_{DE}$ is bounded with a sigmoid function, such
as $w_{DE}=-1+2\sigma \left( s\left( x\right) \right) $, in which $\sigma
\left( s\left( x\right) \right) =\frac{1}{1+e^{-s\left( x\right) }}$ and $%
s\left( x\right) $ is expressed by the Chebyshev polynomials as before.

Last but not least, in order to avoid unphysical oscillations related to
overfitting, we introduce weak smoothness penalties $L_{\mathrm{SP}}$, by
introducing penalties within the loss function on the first and second
derivatives of the trained functions. In particular%
\begin{equation}
\mathcal{L}_{\mathrm{SP}}=\lambda _{\mathrm{SP}}^{\prime }\left( \frac{dF}{dx%
}\right) ^{2}+\lambda _{\mathrm{SP}}^{^{\prime \prime }}\left( \frac{d^{2}F}{%
dx^{2}}\right) ^{2},
\end{equation}%
where $F$\ represent the reconstructed functions. In this study, we assume
that $\lambda _{\mathrm{SP}}^{\prime }$\ and $\lambda _{\mathrm{SP}%
}^{^{\prime \prime }}$\ are weak that is, they have values $10^{-4}$.

\subsection{Cosmological Likelihoods and Loss Function}

The analysis performed in this work is based on late-time cosmological
observations. Specifically, we employ observational datasets from Baryon
Acoustic Oscillations BAO, Type Ia Supernova (SNIa) and Cosmic Chronometers
(CC).

\subsubsection{Cosmic Chronometers}

We employ the cosmic chronometers, which are old galaxies, passively
evolving with synchronous stellar populations and similar cosmic evolution 
\cite{co01}. We consider the model independent direct measurements of the
Hubble parameter within the redshift range $0.09\leq z\leq 1.965$~\cite{co02}%
. The Hubble function is trained by the PINN and the $\chi _{CC}^{2}$ is
defined as%
\begin{equation}
\chi _{\mathrm{CC}}^{2}=\left( H^{\mathrm{PINN}}-H^{obs2}\right)
^{T}C_{CC}^{-1}\left( H^{\mathrm{PINN}}-H^{obs2}\right) ,
\end{equation}%
where the complete covariance matrix\footnote{%
The covariance matrix can be found in
https://gitlab.com/mmoresco/CCcovariance .} is used \cite{co01}

\subsubsection{Baryon Acoustic Oscillations}

We consider the recent BAO DESI DR2 catalogue \cite{desi1,desi2}. This
dataset consists of 13 measurements from DESI DR2 at seven redshifts with
the range $z\in \lbrack 0.295,2.33],$ of the Hubble distance ratio the
comoving angular distance ratio\ and the volume-averaged distance ratio,
each normalized by the sound horizon at the baryon drag epoch $r_{\mathrm{%
drag}}$,%
\begin{equation}
\frac{D_{H}(z)}{r_{\mathrm{drag}}}=\frac{c}{H(z)},\qquad \frac{D_{M}(z)}{r_{%
\mathrm{drag}}}=c\int_{0}^{z}\frac{dz^{\prime }}{H(z^{\prime })},\qquad 
\frac{D_{V}(z)}{r_{\mathrm{drag}}}=\left[ \frac{c\,z\,D_{M}^{2}(z)}{H(z)}%
\right] ^{1/3}.
\end{equation}%
We employ the full $13\times 13$ covariance matrix $\mathbf{C}_{\mathrm{BAO}%
} $, such that the likelihood is defined as

\begin{equation}
\chi _{\mathrm{BAO}}^{2}=(\mathbf{d}_{\mathrm{BAO}}^{\mathrm{PINN}}-\mathbf{d%
}_{\mathrm{BAO}}^{\mathrm{obs}})^{T}\,\mathbf{C}_{\mathrm{BAO}}^{-1}\,(%
\mathbf{d}_{\mathrm{BAO}}^{\mathrm{PINN}}-\mathbf{d}_{\mathrm{BAO}}^{\mathrm{%
obs}}).
\end{equation}%
The comoving distance $D_{M}$ is evaluated by differentiable trapezoidal
quadrature on a grid of $N_{\mathrm{grid}}^{\mathrm{BAO}}$ points.

\subsubsection{Type Ia Supernovae}

We make use of three different (SNIa) compilations, the PantheonPlus (PP)
catalogue without the SH0ES Cepheid calibration \cite{pp1}, the Union3.0
(U3) \cite{union3} and the recent DES-Dovekie (DESD) \cite{desd}. \ The
catalogues provide measurements of the distance modulus $\mu _{i}^{\mathrm{%
obs}}$ as a function of redshift $z_{i}$. The U3 and PP catalogues provide
events within the redshift range $10^{-3}<z<2.27$. The latter catalogues
share 1363 common SNIa events; they are constructed from different
photometric reduction pipelines. On the other hand, the DESD catalogue
released recently after the re-analysis of the five-year Dark Energy Survey
supernova program, yielding 1820 SNIa measurements at lower redshifts $%
z<1.13 $.

The SNIa likelihood marginalizes analytically over the absolute magnitude $M$%
, giving the marginalized $\chi _{\mathrm{SNI}_{a}}^{2}$

\begin{equation}
\chi _{\mathrm{SNI}_{a}}^{2}=\Delta \mu ^{T}\,\mathbf{C}_{\mathrm{SNI}%
_{a}}^{-1}\,\Delta \mu -\frac{(\mathbf{1}^{T}\mathbf{C}_{\mathrm{SNI}%
_{a}}^{-1}\Delta \mu )^{2}}{\mathbf{1}^{T}\mathbf{C}_{\mathrm{SNI}_{a}}^{-1}%
\mathbf{1}},
\end{equation}%
where $\Delta \mu _{i}=\mu _{i}^{\mathrm{obs}}-\mu _{i}^{\mathrm{PINN}}$ and 
$\mu ^{\mathrm{PINN}}(z)=5\log _{10}[(1+z)D_{M}(z)]+25$.

\subsubsection{Data Loss function}

The component of the loss function related to the training given by the
cosmological data is defined as%
\begin{equation}
\mathcal{L}_{\mathrm{DATA}}=\lambda _{\mathrm{CC}}\,\mathcal{L}_{\mathrm{CC}%
}+\lambda _{\mathrm{BAO}}\,\mathcal{L}_{\mathrm{BAO}}+\lambda _{\mathrm{SN}%
}\,\mathcal{L}_{\mathrm{SN}},
\end{equation}%
where $\lambda $ are the data weights, the loss function related to the CC
is defined as 
\begin{equation}
\mathcal{L}_{\mathrm{CC}}=\frac{1}{N_{\mathrm{CC}}}\chi _{\mathrm{CC}}^{2},
\end{equation}%
and for the loss function $\mathcal{L}_{\mathrm{BAO}},~\mathcal{L}_{\mathrm{%
SN}}$ \ we introduce a logarithmic cap, that is, 
\begin{eqnarray}
\mathcal{L}_{\mathrm{BAO}} &=&\gamma ^{2}\ln \!\left( 1+\frac{1}{\gamma ^{2}}%
\frac{\chi _{\mathrm{BAO}}^{2}}{N_{\mathrm{BAO}}}\right) ,  \notag \\
\mathcal{L}_{\mathrm{SN}} &=&\gamma ^{2}\ln \!\left( 1+\frac{1}{\gamma ^{2}}%
\frac{\chi _{\mathrm{SNI}_{a}}^{2}}{N_{\mathrm{SNI}_{a}}}\right) ,
\end{eqnarray}%
where $\gamma $ is a normalized parameter. For $\frac{1}{N_{\mathrm{Data}}}%
\chi _{\mathrm{Data}}^{2}<<\gamma ^{2}$, the $\mathcal{L}_{\mathrm{DATA}%
}\simeq \frac{1}{N_{\mathrm{Data}}}\chi _{\mathrm{Data}}^{2}$. The
logarithmic function has been introduced to avoid a dominance of the high
values of $\chi _{\mathrm{Data}}^{2}$ at the initial states of the training,
when $\frac{1}{N_{\mathrm{Data}}}\chi _{\mathrm{Data}}^{2}>>\gamma ^{2}$. \
Parameter $\gamma ^{2}$ is important for the rescaling of the loss function,
nevertheless, we assume that $\gamma ^{2}>1$.

It is important to note that instead of introducing the standard total $\chi
_{tot}^{2}$ in the loss function we introduce the reduced $\chi
^{2}/N_{data} $ for each observational data set. We made this selection in
order to avoid the dominance of the datasets with the larger number of data,
i.e. the SNIa catalogue. Thus, with this approach we ensure a balanced
contribution among the different cosmological data during the reconstruction
process. This procedure modifies the relative statistical weighting of the
datasets and therefore should be interpreted as a reconstruction criterion
and not as a strict likelihood combination.

The data loss components are presented within the yellow box in Fig. \ref%
{fig0}.

\subsection{Total Loss Function}

The total loss function of Cosmo-PINN is constructed as a weighted
combination of all the individual components introduced above, that is, 
\begin{equation}
\mathcal{L}_{\mathrm{PINN}}=\mathcal{L}_{\mathrm{PDE}}+\mathcal{L}_{\mathrm{%
DATA}}+\mathcal{L}_{\mathrm{IC}}+\mathcal{L}_{\mathrm{SP}}+\mathcal{L}_{%
\mathrm{bounds}}.
\end{equation}%
The weights for the PDE loss component and of the $\mathcal{L}_{\mathrm{DATA}%
}$, that is, $\lambda _{\mathrm{PDE}}$, $\lambda _{\mathrm{CC}}$, $\lambda _{%
\mathrm{BAO}}$ and $\lambda _{\mathrm{SN}}$ are adapted dynamically during
training using the GradNorm method introduced in \cite{gradnorm}. We
consider adaptive weights because the different components of the loss
function have different scales during training, and by considering constant
weights it is possible for one of the loss components to dominate over the
others.

\subsection{Training}

The network consists of four hidden layers with 64 neurons each, and as
activation function we consider the $\tanh $. The training proceeds in two
sequential phases, an Adam optimizer and the L-BFGS optimizer epochs. The
Adam optimizer is employed for 10000 epochs and the L-BFGS optimizer is
applied for up to 200 epochs, with early stopping when the total loss
reaches a minimum. The dynamic adjustment of the loss weights$~\lambda _{%
\mathrm{PDE}}$, $\lambda _{\mathrm{CC}}$, $\lambda _{\mathrm{BAO}}$ and $%
\lambda _{\mathrm{SN}}${} via the GradNorm method take place for the first
5000 epochs of the Adam optimizer. After this epoch the weights are frozen
and remain fixed for the rest of the training.

\subsection{Posterior Uncertainties}

In order to determine the uncertainties of the reconstructed functions, that
is, $\bar{\rho}_{DE}(z)$, $H(z)$ and $w_{DE}(z)$, we use the solution
obtained from the L-BFGS optimizer. We employ the Hamiltonian Monte Carlo%
\footnote{%
https://github.com/vrettasm/HamiltonianMonteCarlo} (HMC) \cite{hmc1}
approach with the NUTS sampler \cite{hmc2}, sampling the posterior
distribution over the free cosmological parameters $\Omega _{m0}$, $H_{0}$
and $r_{\mathrm{drag}}$, together with the Chebyshev coefficients that
parametrize $w_{DE}(z)$ and the weights of the final network layer that
parametrize $\bar{\rho}_{DE}(z)$. The Gaussian priors defined before for the
cosmological parameters are introduced, without the corresponding weights.
Moreover, the log-likelihood is given by the combined contribution of the
cosmological data in addition with the residual which describes the physics.
We run four independent chains, initialized by over-dispersing the trained
solution, such that the Gelman-Rubin convergence criterion satisfies $\hat{R}%
<1.01$ for all the trained variables. Propagating the posterior samples
through the network we obtain the $68\%$ and $95\%$ credible intervals on
the reconstructed functions. At this point, it is important to mention that
in contrast to the training loss, the HMC likelihood uses the standard total 
$\chi _{\min }^{2}$\ of each dataset, without the reduced $\chi _{\min
}^{2}/N_{\mathrm{data}}$\ normalization or the logarithmic cap. The
statistical weights of the datasets are therefore preserved in the
uncertainty quantification.

Before we proceed with the reconstruction of the dark energy equation of
state parameter $w_{DE}\left( z\right) ~$we perform two tests to assess the
viability of the Cosmo-PINN. First, we perform training runs with different
sets of initial conditions to verify whether the PINN converges to a
consistent solution. Second, we train the network using expansions of
different orders of Chebyshev polynomials to study their effect on the
obtained solutions, as also we train the model by considering $w_{DE}\left(
z\right) $\ to be described by a network of two hidden layers with 32
neurons and we compare the obtained reconstruction with that of Chebyshev
expansion. The results of these tests are summarized in Appendix \ref%
{stabtest}.

For the analysis which follows below, we adopt the Chebyshev expansion
representation and we consider the sixth-degree Chebyshev polynomial, with
only seven additional coefficients instead\ of the $n\times m$ free
parameters of the second network, as it provides a computationally less
demanding representation.\ The reconstruction of the physical dark energy
function is independent of the reparametrization approach for the $%
w_{DE}\left( z\right) $.

\section{Reconstruction of the Dark Energy Equation of State}

\label{sec4}

We have demonstrated that the network is stable and capable of training
solutions which simultaneously satisfy the observational data and the
corresponding physical laws. We proceed with the reconstruction of the dark
energy equation of state parameter $w_{DE}\left( z\right) $. For the
training we select soft prior constraints for the cosmological parameters $%
H_{0},~\Omega _{m0}$ and $r_{\mathrm{drag}}$, near to the Planck values.
From expression (\ref{c.16}) it follows that $w_{DE}\left( z\right) $ and $%
\Omega _{m0}$ are not independent. Therefore the introduction of a prior on $%
\Omega _{m0}$ is necessary in order to avoid degeneracies in the
reconstructed solution. On the other hand, if we assume a smaller value for
the $\Omega _{m0}$, the Cosmo-PINN will reconstruct a unified dark
energy-dark matter cosmological scenario.

\begin{figure}[tbph]
\centering\includegraphics[width=1\textwidth]{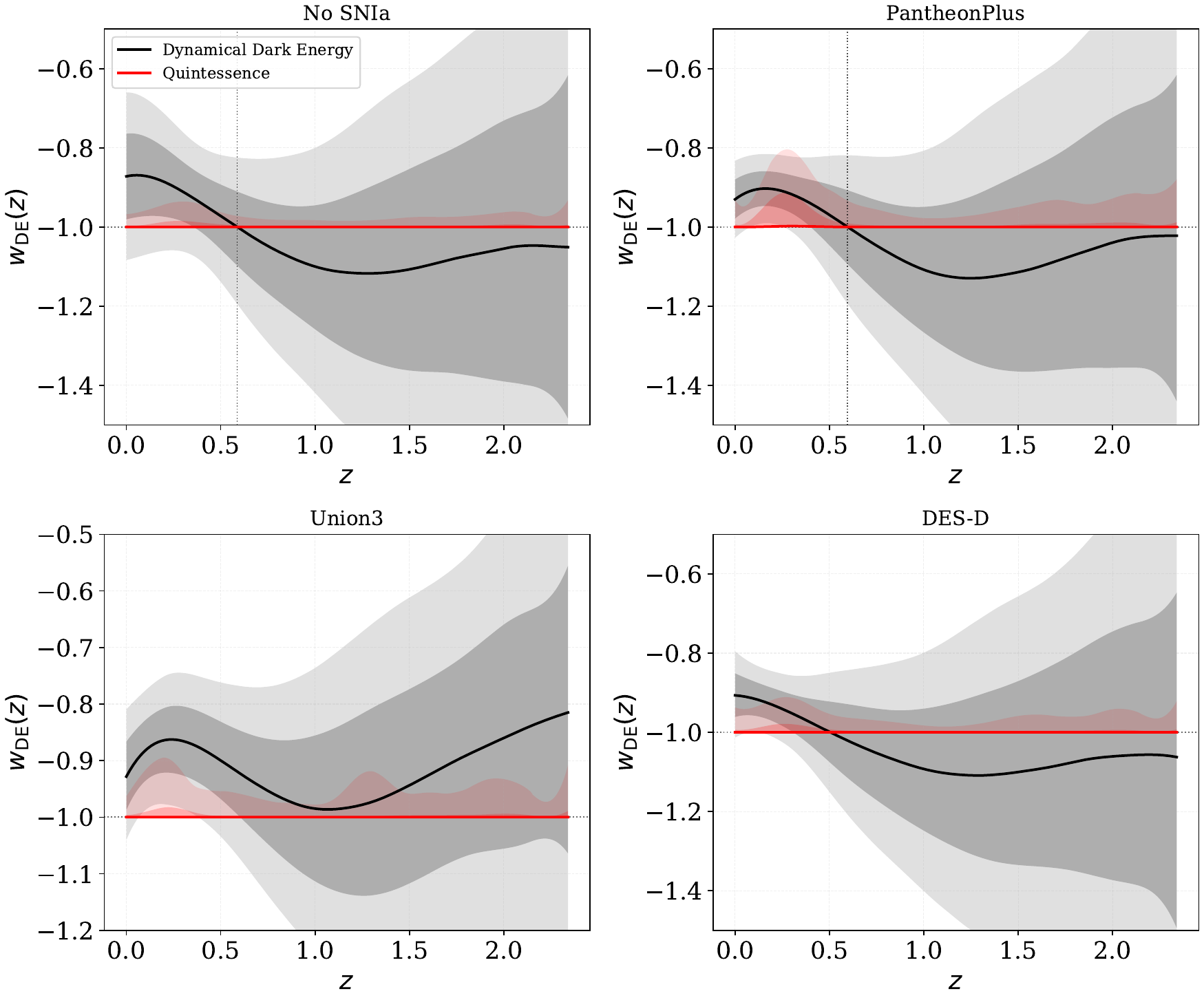}
\caption{Reconstruction of the dark energy equation of state parameter $%
w_{DE}\left( z\right) $ for four different data set combinations. Black
lines are for the case where $w_{DE}\left( z\right) $ is allowed to cross
the phantom divide line, and red lines are for the quintessence case. Bands
are for the $68\%$ and $95\%$ credible intervals.}
\label{fig6}
\end{figure}

We reconstruct the dark energy equation of state parameter for four
different combinations of the observational datasets. Specifically, we
consider the combinations CC+BAO and CC+BAO+SNIa, where the SNIa dataset
corresponds to the PP, U3 and DD compilations. The reconstruction is
performed under the strong physical constraint $\Omega _{DE}\left( z\right)
\in \left[ 0,1\right] $. Furthermore, for the dark energy equation of state
parameter $w_{DE}\left( z\right) $ we investigate two different scenarios,
in the first case $w_{DE}\left( z\right) $ can cross the phantom divide
line, and $w_{DE}\left( z\right) $ has the value $-1$ as lower bound, such
that $w_{DE}\left( z\right) \geq -1$.

\begin{figure}[h]
\centering\includegraphics[width=1\textwidth]{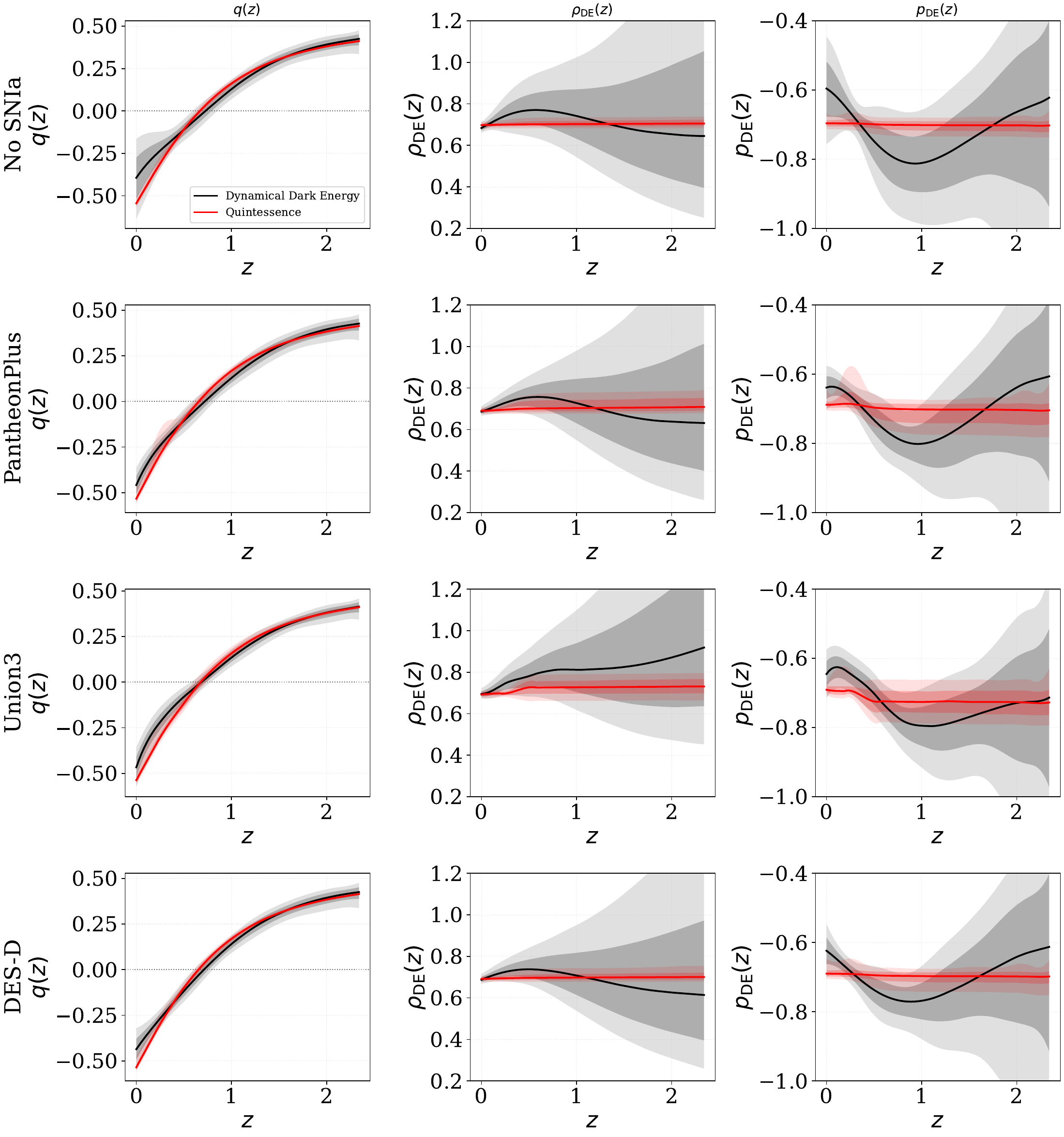}
\caption{Reconstruction of the cosmological parameter $q\left( z\right) $,~$%
\Omega _{m}\left( z\right) ~$(solid lines)$,~\Omega _{DE}\left( z\right) $
(dashed lines) and $p_{DE}\left( z\right) $ for the four different
combinations of the data sets. Black lines are for the case where $%
w_{DE}\left( z\right) $ is allowed to cross the phantom divide line, and red
lines are for the quintessence case. Bands are for the $68\%$ and $95\%$
credible intervals.}
\label{fig7}
\end{figure}

\begin{figure}[h]
\centering\includegraphics[width=1\textwidth]{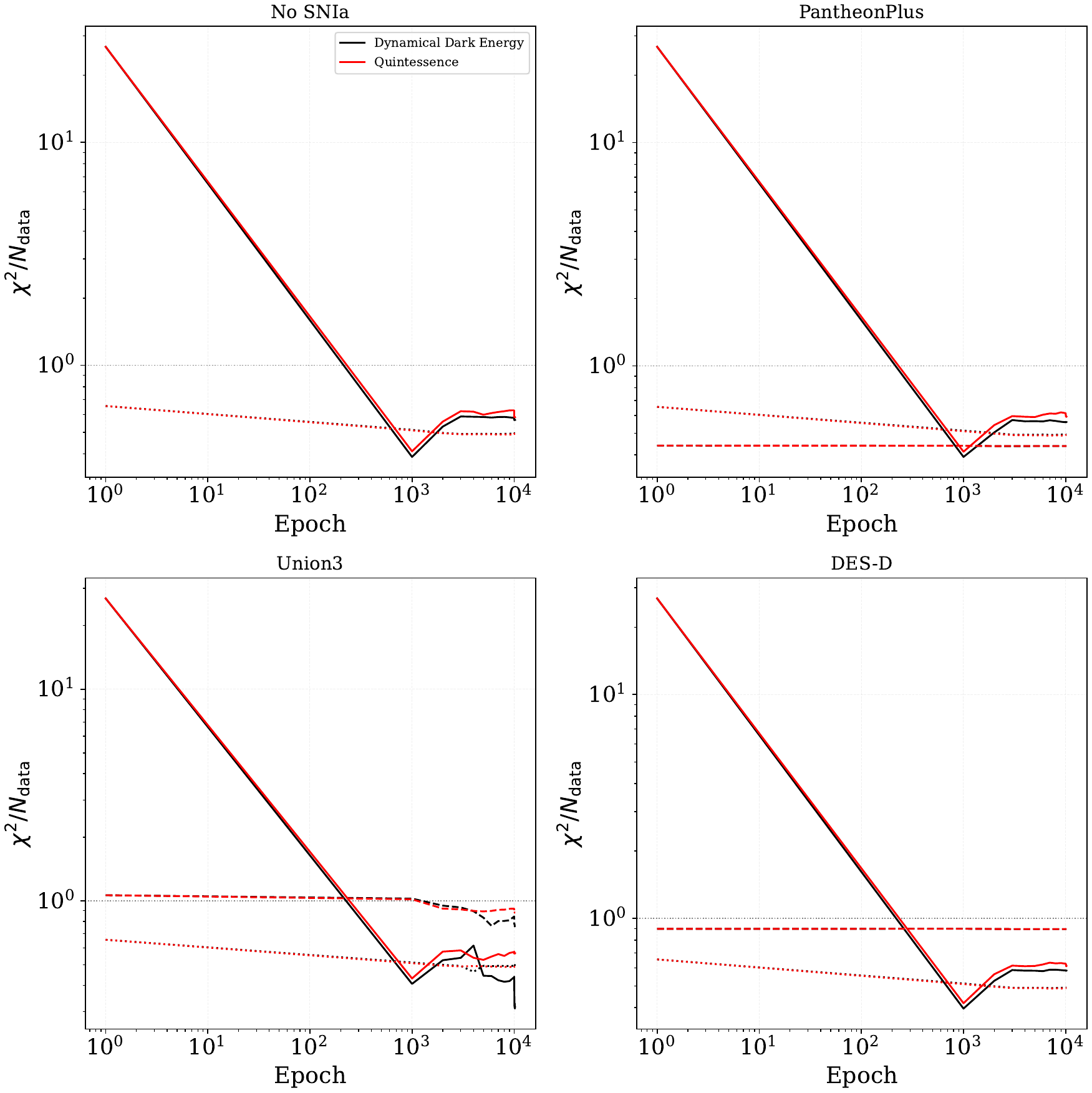}
\caption{Evolution of the $\protect\chi ^{2}/N_{data}$ \ for each
observable~during training for the four different dataset combinations.
Black lines are for the dynamical dark energy model and red lines for the
quintessence. }
\label{fig10}
\end{figure}

\begin{figure}[h]
\centering\includegraphics[width=1\textwidth]{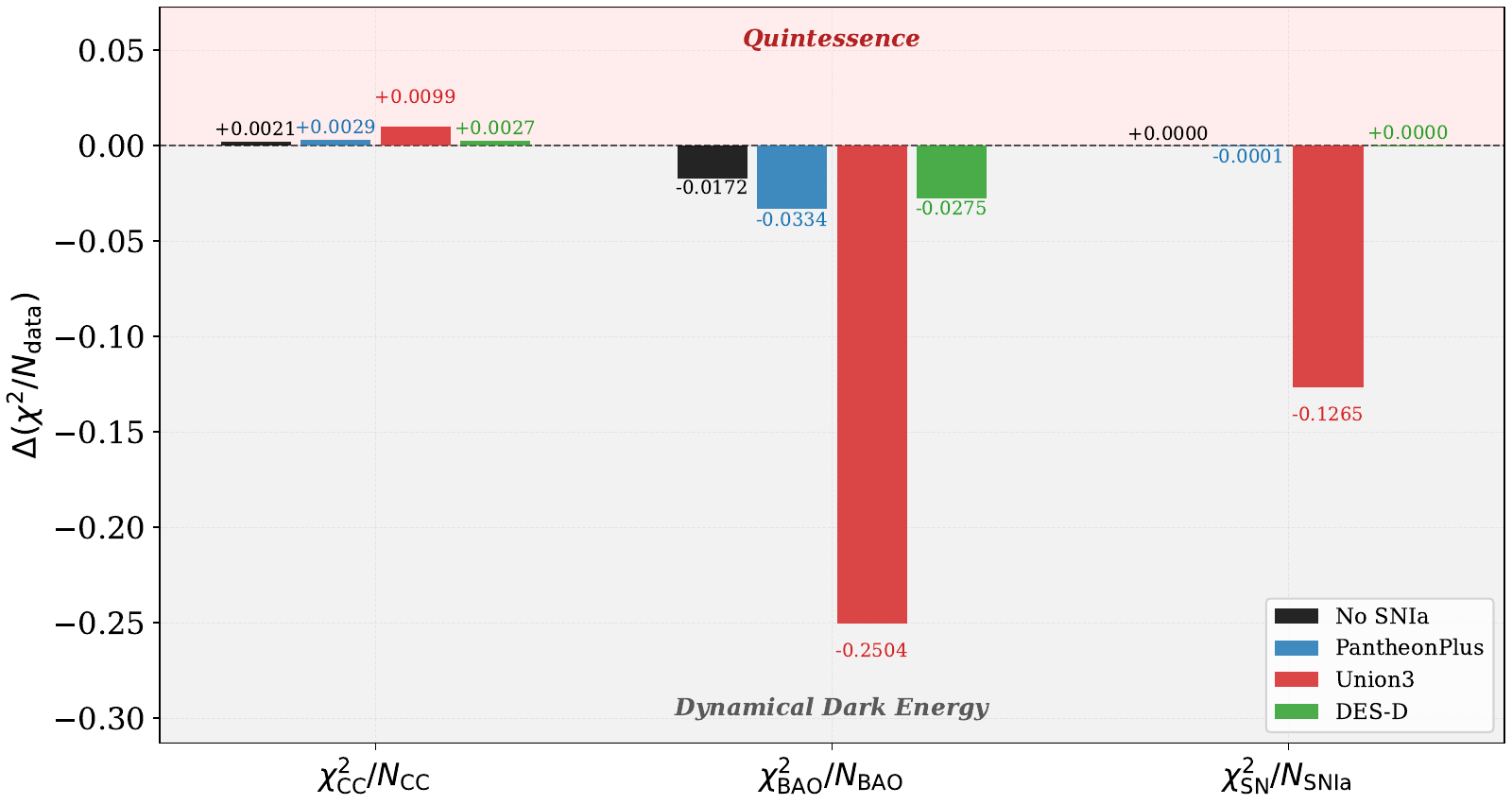}
\caption{Differences in $\protect\chi ^{2}/N_{data}$ ~training diagnostic
between the dynamical dark energy and quintessence reconstructions, for the
four dataset combinations. Negative values indicate that the dynamical dark
energy reconstruction attains a lower $\protect\chi ^{2}/N_{data}$ \
positive values indicate the same for the quintessence reconstruction. These
differences are optimization diagnostics only and should not be interpreted
as likelihood ratios.}
\label{fig9}
\end{figure}

In Fig. \ref{fig6} we present the reconstructed $w_{DE}\left( z\right) $,
while in Fig. \ref{fig7} the evolution of the physical parameters $q\left(
z\right) $,~$\bar{\rho}_{DE}\left( z\right) $ and $p_{DE}\left( z\right) $
are given for the four different datasets applied. When the $w_{DE}\left(
z\right) $ has no lower bound, that is, the corresponding weight in the loss
function is zero, i.e. $\lambda _{wDE\geq -1}=0$, the $w_{DE}\left( z\right) 
$ at the present $w_{DE}\left( 0\right) >-1,$ while it crosses the phantom
divide line near $z=0.5$. The function is monotonically decreasing. The
latter are in agreement with the results obtained by the CPL model \cite%
{anlcdm}. The corresponding physical quantities have a similar behaviour
independent of the SNIa compilation, where the acceleration starts around $%
z\simeq 0.7$. However, the $\Omega _{DE}\left( z\right) \rightarrow 0$, such
that the effects of the dark energy fluid in the past are negligible.

On the other hand, in the quintessence scenario, the behaviour of $%
w_{DE}\left( z\right) ~$exhibits a different evolution. In all the data
combinations the reconstructed function $\bar{\rho}_{DE}\left( z\right) $ is
forced to be constant, that is, the equation of state parameter reaches the $%
\Lambda $CDM limit, i.e. $w_{DE}\left( z\right) \rightarrow -1.~$%
Consequently, the effective dark energy component behaves almost as a
cosmological constant, with negligible deviations from the $\Lambda $CDM
scenario. Last but not least in the quintessence scenario, the $\Omega _{m0}$
is trained to have a much lower value from the prior considered, indicating
that the soft constraints introduced in $\mathcal{L}_{\mathrm{IC}}$ does not
dominate over the reconstructed solutions.

The evolution of the $\chi ^{2}/N_{data}$ during training is presented in
Fig. \ref{fig10} for each observable for the four different datasets. We
observe that the training history is similar for the quintessence and the
dynamical dark energy model. There are small differences in the first 1000
epochs of the training. Fig. \ref{fig9} presents the difference in $\chi
^{2}/N_{data}$ ~between the dynamical dark energy and the quintessence at
the reconstructions.

The SNIa data are reproduced with comparable fit quality by both
reconstructions, with a marginal difference for the U3 catalogue in favour
of the dynamical dark energy case. The CC observable shows have a small
support over the quintessence scenarios. For the BAO data, a lower $\chi
^{2}/N_{BAO}$ for the dynamical dark energy reconstruction is observed only
for all the data combinations, with a strong support when the U3 SNIa data
are considered.

\begin{figure}[h]
\centering\includegraphics[width=1\textwidth]{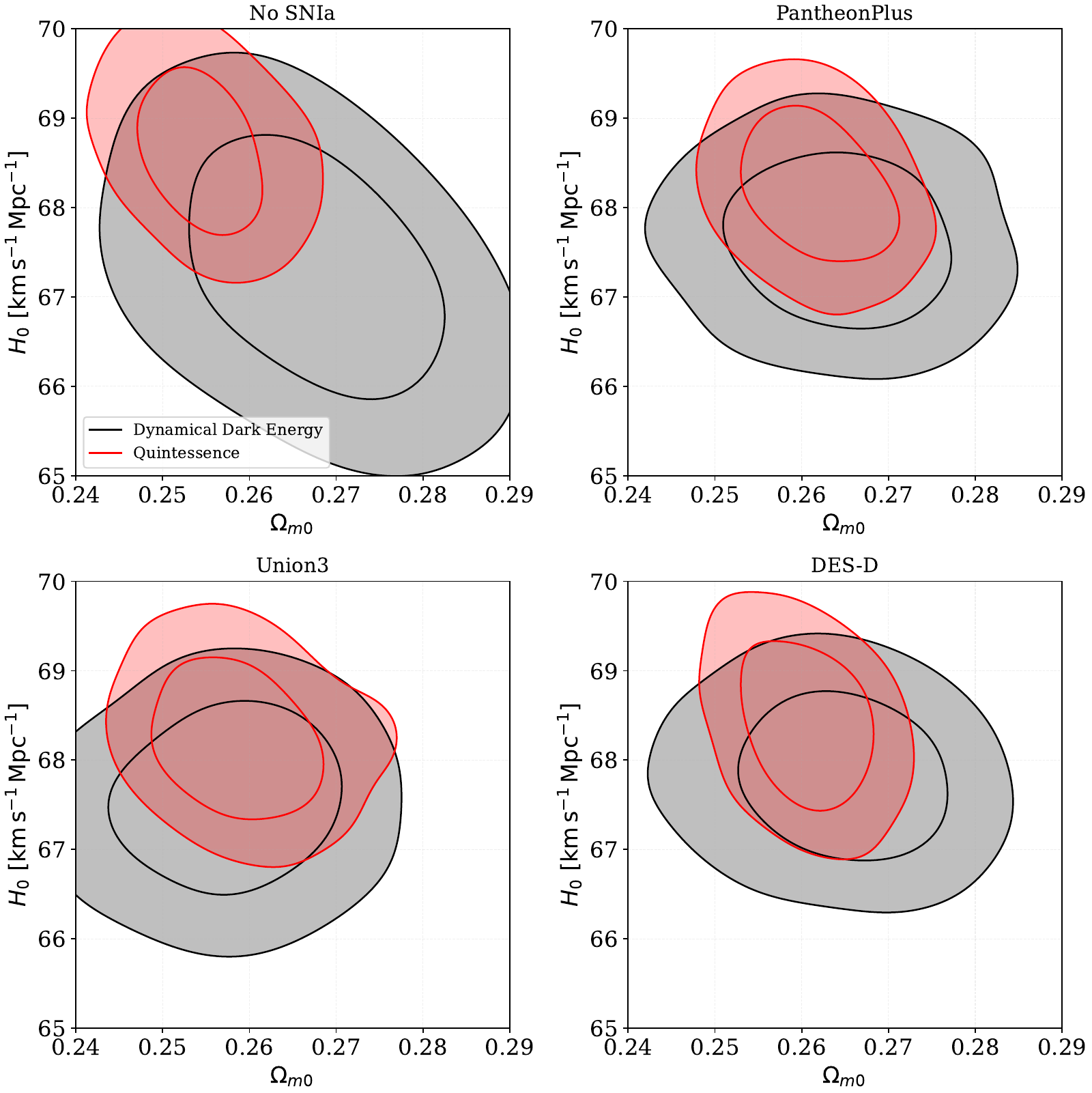}
\caption{Confidence contours for the trained parameters $H_{0}$ and $\Omega
_{m0}$ for the four different datasets, for the dynamical dark energy and
the quintessence reconstructions, as follows from the HMC approach.}
\label{fig12}
\end{figure}

Finally, in Fig. \ref{fig12} we present the confidence contours of the
trained cosmological parameters $H_{0}$ and $\Omega _{m0}$ as obtained from
the analysis of the four chains obtained by the HMC approach. 

\section{The Role of Physical Constraints within the Neural Network}

\label{sec5}

\begin{figure}[t]
\centering\includegraphics[width=1\textwidth]{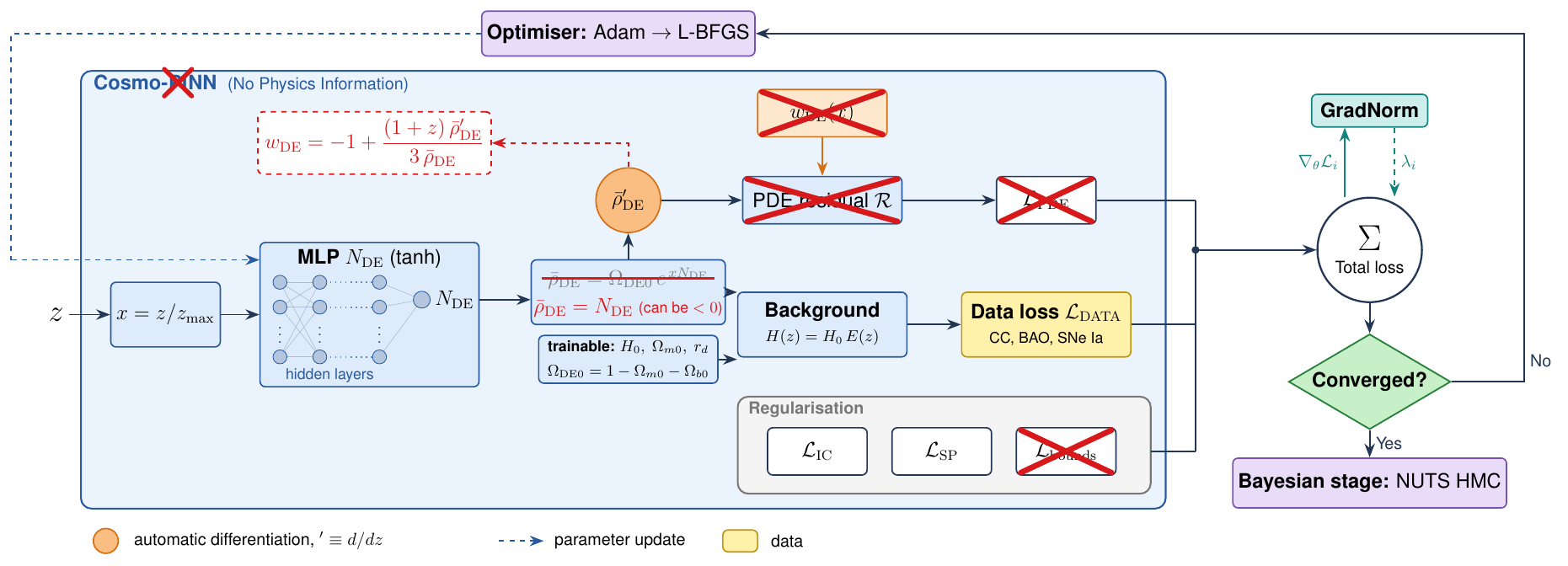}
\caption{Architecture of the NN without Physical Information. The components
of the Cosmo-PINN that are removed are crossed out, they are the exponential
ansatz for the dark energy density, the physics residual, and the physically
bounds terms of the loss function. The dark energy equation of state
parameter is derived directly from reconstructed dark energy density.}
\label{figar}
\end{figure}

In order to understand the importance of the physical law and of the
physical constraints within the NN during the reconstruction we perform the
same reconstruction approach by using the CC+BAO+PP dataset. We consider a
NN where the trained physical variable is the Hubble function as obtained
directly by the cosmological data. After the reconstruction of the $H\left(
z\right) $ the physical parameters $w_{DE}\left( z\right) ,~\Omega
_{m}\left( z\right) $ and $\Omega _{DE}\left( z\right) $ are derived. The
two networks share the same initial conditions and the same network
architecture. For the NN, the $\rho _{DE}\left( z\right) $ is defined as $%
\rho _{DE}\left( x\right) =N_{DE}\left( x\right) $ without the exponential
function, this means that $\rho _{DE}\left( x\right) <0$ is allowed\footnote{%
We remark that recently there exists an interest in models where the dark
energy density changes sign, such as the sign-switching cosmological
constant model \cite{occ1,occ2}. For the description of this model, a
definition for the $\rho _{DE}\left( z\right) $ will be $\rho _{DE}\left(
z\right) =3H_{0}^{2}\left( \Omega _{DE}+xN_{DE}\left( z\right) \right) $.}.
Furthermore, we impose the same smoothness penalties with the same weight.
Another difference between the two networks is that the $\mathcal{L}_{%
\mathrm{PDE}}$ is not introduced in the total loss function. The basic
architecture of the NN and the differences with the Cosmo-PINN are
demonstrated in Fig. \ref{figar}.

\begin{figure}[h]
\centering\includegraphics[width=1\textwidth]{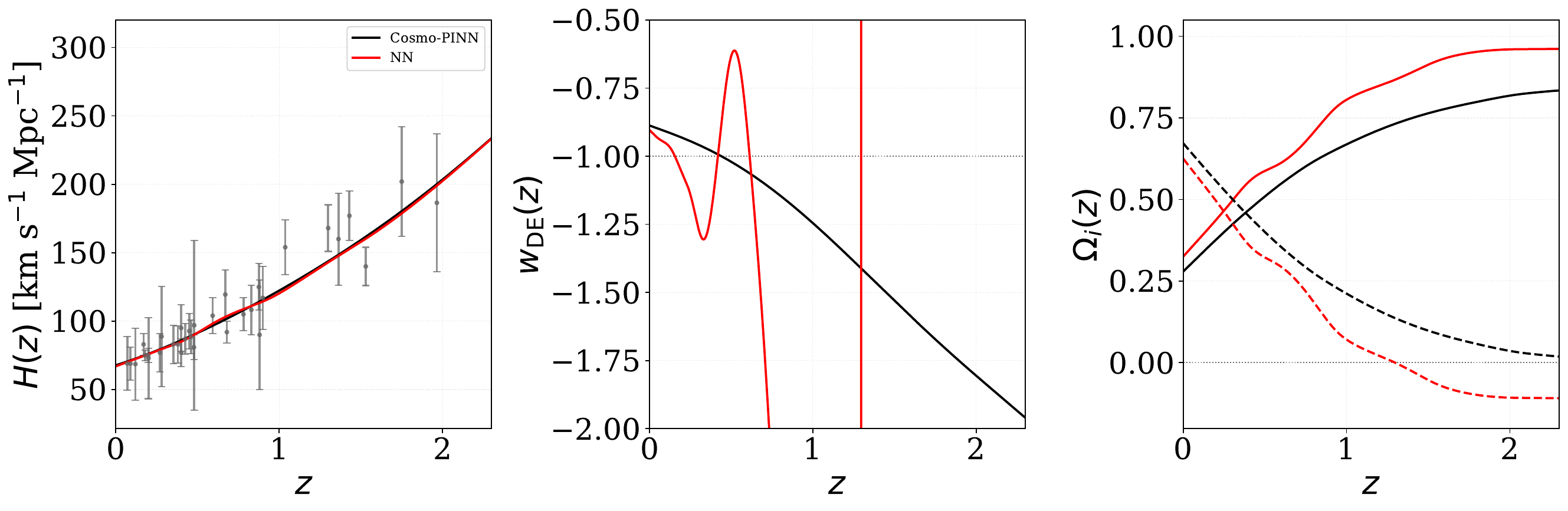}
\caption{Comparison of the reconstructed cosmological parameters between the
Cosmo-PINN and the NN with the same architecture, without the physical
constraints in the loss function. }
\label{fig8}
\end{figure}

The comparison of the cosmological parameters of the NN with those obtained
from the Cosmo-PINN is presented in Fig. \ref{fig8}. The Hubble function
reconstructed by the NN is similar to that obtained from the Cosmo-PINN,
that is, an expected result because in both networks the Hubble function is
fitted to the data. Small oscillations appear in the NN reconstruction,
which can be attributed to the absence of both the Chebyshev expansion and
the physical constraints that are imposed on the solution in the Cosmo-PINN.
Moreover, the obtained reconstructed solution for the $\Omega _{DE}\left(
z\right) $, becomes negative as we reach the matter epoch, which explains
the ill-defined behaviour of the $w_{DE}\left( z\right) $ during the
transition epoch. Finally, in the low-redshift regime oscillations are also
present in $w_{DE}\left( z\right) ,$ that is largely a numerical artifact
due to the direct derivation from the $H\left( z\right) $, and not through
the training of the network as in\ the case of the Cosmo-PINN.

\section{Conclusions}

\label{sec6}

We have introduced Cosmo-PINN, a framework for the reconstruction of free
functions and parameters of cosmological theories. In Cosmo-PINN, the
physical laws are embedded directly within the loss function of the network,
ensuring that the trained solution satisfies the corresponding field
equations at every point of the redshift region where the reconstruction
takes place.

We focus on the reconstruction of the dark energy equation of state
parameter $w_{DE}\left( z\right) $, by using late-time observational data,
such as the BAO measurements from DESI DR2, the CC and three catalogues for
the SNIa, the PP, the U3 and the DD. Furthermore, the cosmological
parameters $H_{0},~\Omega _{m0}$ and $r_{\mathrm{drag}}$, \ are treated as
training variables, constrained by soft priors anchored near the Planck 2018
values. The framework is trained to reconstruct the Hubble function $H\left(
z\right) $ from the cosmological data, while the $\bar{\rho}_{DE}\left(
z\right) $ and the $w_{DE}\left( z\right) $ are recovered through the
cosmological field equations embedded in the loss function. Smoothness
penalties and physical boundary conditions are additionally imposed to
ensure the lack of numerical artifacts in the reconstructed solution. The
equation of state parameter $w_{DE}\left( z\right) $\ is represented either
by a second neural network or by a Chebyshev polynomial expansion, which
avoids overfitting and numerical oscillations.

Furthermore, two main tests have been performed in order to study the
stability and the robustness of the network. Specifically, we studied the
reconstruction process by using different degrees of the Chebyshev
polynomials and of the network reparametrization for the $w_{DE}\left(
z\right) $. We found that for $N\geq 5$, the obtained results are
statistically equivalent. The neural network representation of $w_{DE}\left(
z\right) $ leads to a reconstructed solution which is statistically
indistinguishable from the Chebyshev expansion. Furthermore, in order to
examine the stability of the network we trained the model with different
sets of initial values for the cosmological parameters $H_{0},~\Omega _{m0}$
and $r_{\mathrm{drag}}$. The reconstructed solutions are consistent across
all runs and insensitive to the choice of initial conditions, demonstrating
that the network is stable and that the training converges to a unique
physical solution.

For the reconstruction of the dark energy equation of state parameter $%
w_{DE}\left( z\right) $ we considered two scenarios.\ In the first, $%
w_{DE}\left( z\right) $ is allowed to cross the phantom-divide line, while
in the second the quintessence bound $w_{DE}\left( z\right) \geq -1$ is
imposed in the network as a physical constraint. In the first scenario the $%
w_{DE}\left( z\right) $ is found to be a monotonically decreasing function
that crosses the phantom divide near to $z=0.5$. This behaviour is
consistent with the CPL model. Furthermore, the $\Omega _{DE}\left( z\right) 
$ reaches zero at higher redshifts which means that there is zero
contribution of the dark energy fluid in the cosmological fluid in the
matter dominated era.

In the quintessence scenario the picture changes dramatically, where the $%
w_{DE}\left( z\right) $ is reconstructed to be a monotonically increasing
function for three of the four data set combinations, and when the U3
supernova are employed the $w_{DE}\left( z\right) $ has the local minimum $%
-1.$ The obtained values of the $\Omega _{m0}$ are systematically smaller
than in the unbounded case, and the contribution of the quintessence field
to the total cosmic fluid at high redshifts is found to be non-zero,
revealing a unified mechanism in which the quintessence field mimics the
matter component. This behaviour of the physical parameters is naturally
described by the exponential potential \cite{copeland}, or by a hyperbolic
potential \cite{rubano} when the U3 SNIa data are employed. The explicit
reconstruction of the scalar field potential, however, lies beyond the scope
of the present work and will be addressed in a forthcoming study.

Finally, we consider a neural network with the same architecture, the same
datasets, and the same smoothness penalties as Cosmo-PINN, but without
imposing the cosmological field equations as constraints in the loss
function. The purely data-driven reconstruction produced small oscillations
in the Hubble function leading to numerical artifacts in the reconstructed
physical parameters. Furthermore, the data was found to support a ghost
scenario where the $\Omega _{DE}\left( z\right) $ can have values $\Omega
_{DE}\left( z\right) <0$ during the matter dominated era, which led to an
ill-defined behaviour for the $w_{DE}\left( z\right) $ at the transition
redshift. This comparison demonstrates the importance of imposing the
physical laws at the loss function or within the architecture of the network
to get physically consistent solutions. This shows the differences between
standard ANN and a PINN framework.

The Cosmo-PINN provides a stable and physically consistent framework for the
reconstruction of cosmological observables and of the underlying
cosmological theory. In this study we applied the Cosmo-PINN architecture to
the reconstruction of the dark energy equation of state parameter $%
w_{DE}\left( z\right) $, nevertheless, the framework is general and can be
naturally extended to the reconstruction of free functions in scalar field
cosmologies, in modified theories of gravity, and in interacting dark
energy--dark matter scenarios, where the field equations admit an analogous
PDE-residual formulation. The Cosmo-PINN framework has been shown to extract
physical information directly from the cosmological observations while
remaining consistent with the underlying physical laws.

Posterior uncertainties on the reconstructed functions are obtained via HMC
sampling, and the obtained confidence intervals derived. A systematic study
of the sensitivity of the uncertainty quantification to the choice of
sampler, prior, and network architecture will be performed elsewhere.

The cosmological model considered here, is a generalized parametric dark
energy-dark matter model without interaction and it is integrable, when a
specific function of the equation of state parameter $w_{DE}(z)$\ is given.
Thus, in this scenario, a standard parameter estimation over the free
parameters of the model is sufficient, since the field equations are
self-consistent. Nevertheless, the present approach should not be confused
with the derivation of $w_{DE}\left( z\right) $\ through MCMC analysis. In
the latter, a specific parametric function for the $w_{DE}\left( z\right) $\
is provided which is used to construct the $\rho _{DE}\left( z\right) $\ via
equation (\ref{c.15}) and the corresponding Hubble function. In this
framework the reconstruction follows a different direction where the $\rho
_{DE}\left( z\right) $\ is constrained directly by the data which is the
physical quantity trained by the network. The $w_{DE}\left( z\right) $\ does
not contribute to the data loss terms, and it is derived through the
residual for equation (\ref{c.13}). The Chebyshev expansion is a series
approximation of the solution $w_{DE}\left( z\right) \,$provided by the
trained function for the $\rho _{DE}\left( z\right) \,$and not an exact
parametric form for the dark energy equation of state parameter, as has been
shown from the tests we did with the neural network representation.
Cosmo-PINN provides two statistically equivalent approaches for the
reconstruction of $w_{DE}\left( z\right) $, the neural network
representation and the Chebyshev expansion representation which offers a
smaller number of degrees of freedom.

In the Cosmo-PINN, the physical law is imposed through the residual in the
loss function and is therefore satisfied only to within the accuracy of the
training. Nevertheless, the motivation for the Cosmo-PINN lies in the cases
where the cosmological model is not integrable, or the free functions are
not justified. In these models, the parametric analysis requires assumptions
for the free functions of the model, and numerical techniques for the
derivation of the observable from the field equations. Nevertheless, with
the Cosmo-PINN we can perform the task with a simple training. Consequently,
a natural extension of this framework is to apply it to non integrable
cosmological theories, as also to the level of the linear perturbations.
While this work was under consideration, an independent study appeared in 
\cite{pnn5a} presenting an independent reconstruction of the Hubble function
using the growth of matter perturbations. Nevertheless, the neural network
architecture adopted in that work differs from the one introduced here.

\appendix%

\section{Stability and Robustness Tests}

\label{stabtest}

In this Appendix we present the stability and robustness tests for the
Cosmo-PINN.

\subsection{Initial Condition test}

\begin{figure}[tbph]
\centering\includegraphics[width=1\textwidth]{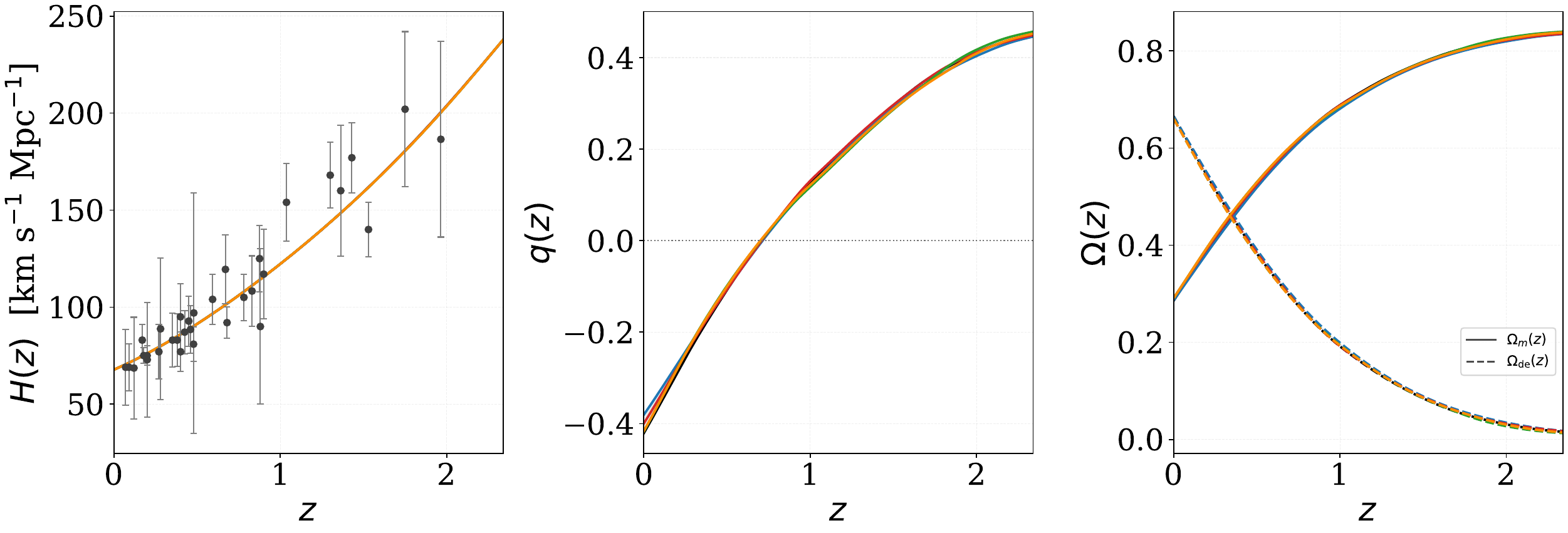}
\caption{Trained Hubble function $H\left( z\right) $, deceleration parameter 
$q\left( z\right) $ and evolution of the energy densities $\Omega _{m}\left(
z\right) $ and $\Omega _{DE}\left( z\right) $, for five different sets of
initial conditions.}
\label{fig1}
\end{figure}

\begin{figure}[htbp]
\centering\includegraphics[width=1\textwidth]{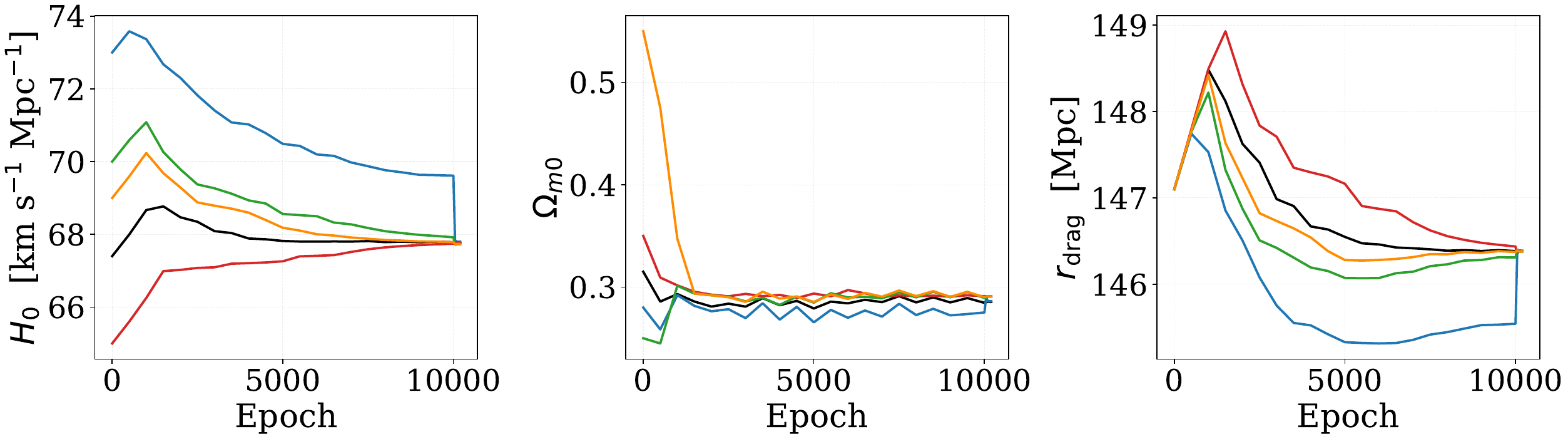}
\caption{Training trajectories of the cosmological parameters $H_{0},~\Omega
_{m0}$ and $r_{\mathrm{drag}}$, for five different set of initial
conditions. The figures indicate that the trajectories reach consistent
values during the training independent of the initial conditions.}
\label{fig2}
\end{figure}

\begin{figure}[htbp]
\centering\includegraphics[width=1  \textwidth]{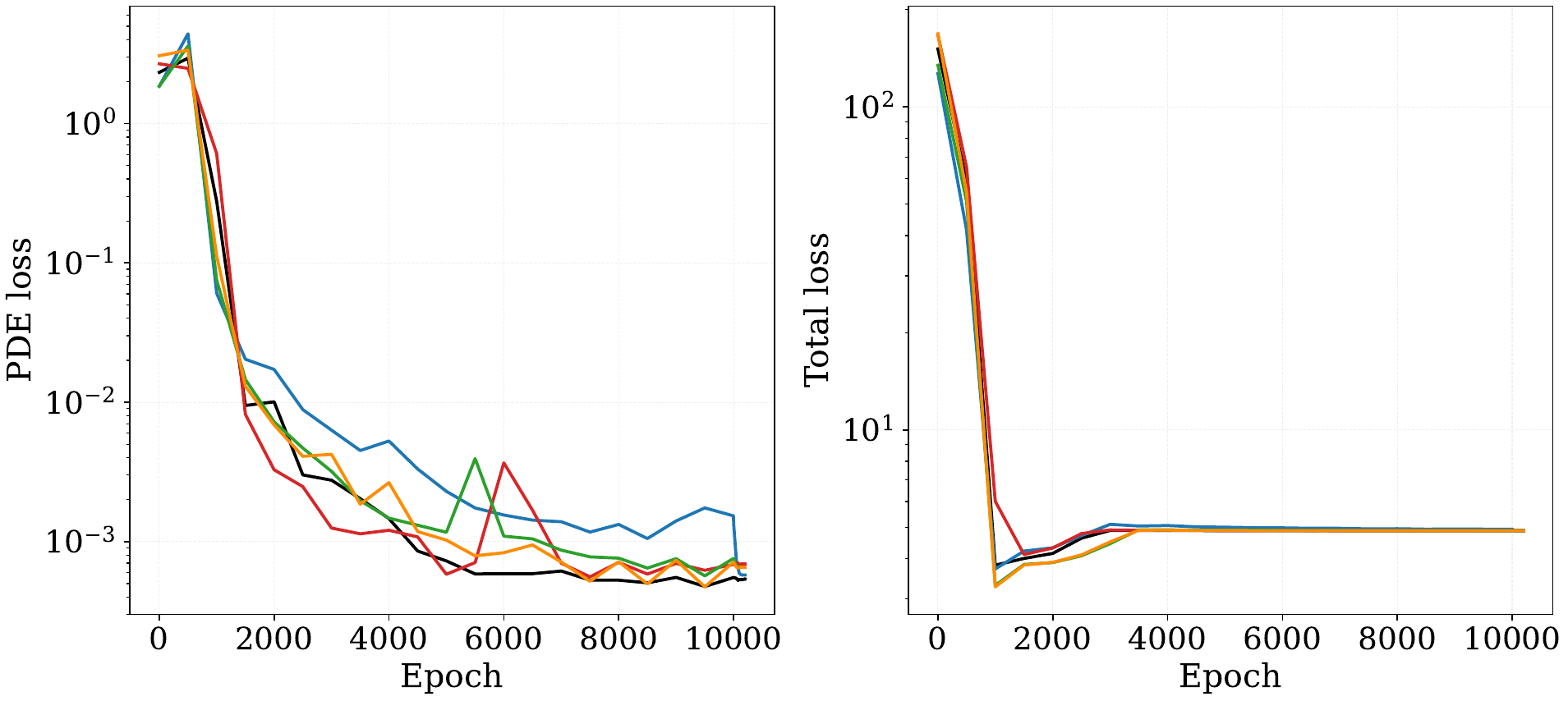}
\caption{Evolution of the PDE loss parameter and of the total loss function
during training for five different sets of initial conditions.}
\label{fig3}
\end{figure}

In Fig. \ref{fig1} we show the reconstructed Hubble function, the
deceleration parameter and the energy densities for the five different
trainings. The training trajectories of the cosmological parameters $H_{0}$, 
$\Omega _{m0}$ and $r_{drag}$ are presented in Fig. \ref{fig2}, while in
Fig. \ref{fig3} we present the evolution of the total loss and of the PDE
loss functions during the training. We note that the solutions provided by
the Cosmo-PINN are consistent and insensitive to the choice of initial
conditions. For these training runs, the prior of the cosmological
parameters was selected to be medium.

\subsection{$w_{DE}\left( z\right) $ Representation}

\begin{figure}[tbph]
\centering\includegraphics[width=1\textwidth]{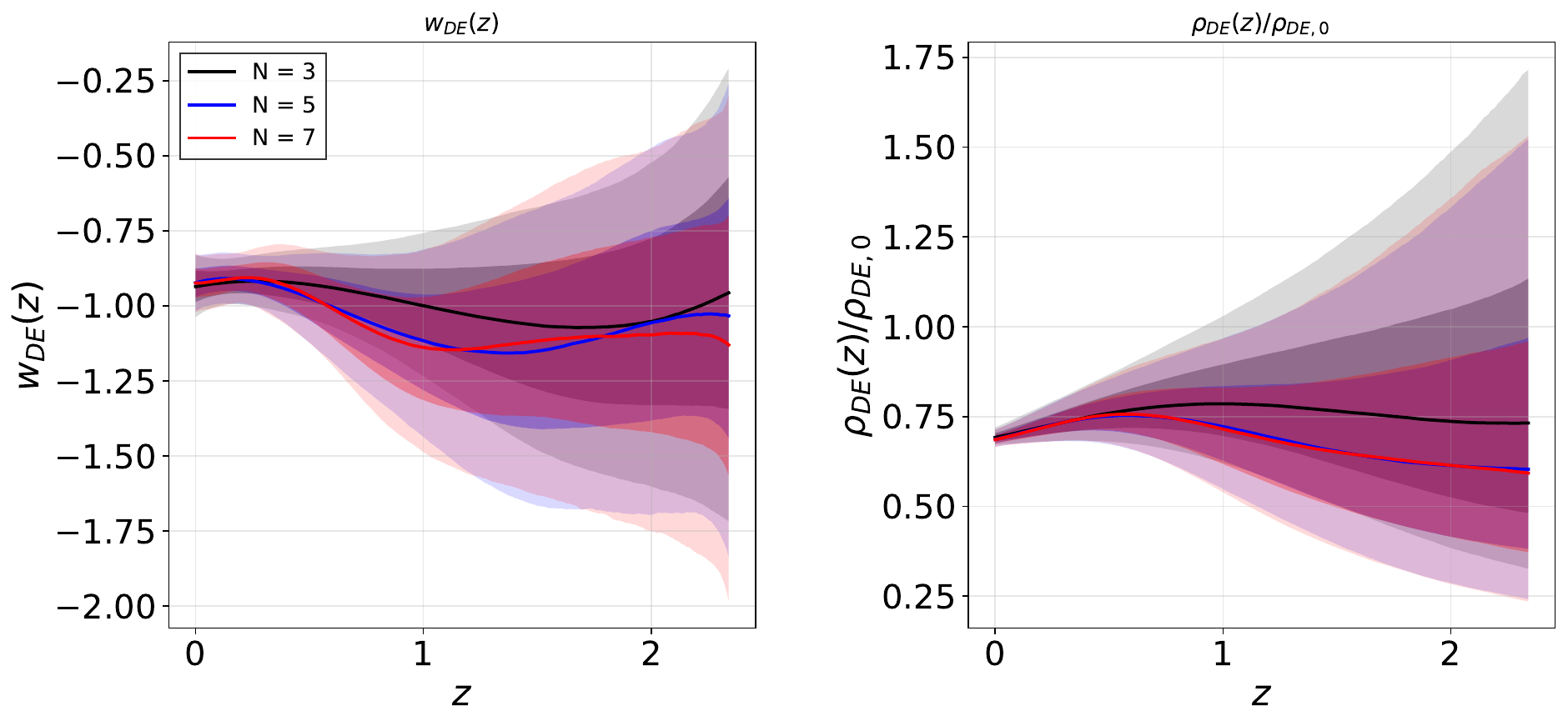}
\caption{Reconstructed dark energy equation of state parameter $w_{DE}\left(
z\right) $ and $\protect\rho _{DE}\left( z\right) $ for different orders of
the Chebyshev polynomial and combined data CC+BAO+PP. The plots are for $%
N=3, $ $5~$and $7$.}
\label{fig4}
\end{figure}

\begin{figure}[tbph]
\centering\includegraphics[width=1\textwidth]{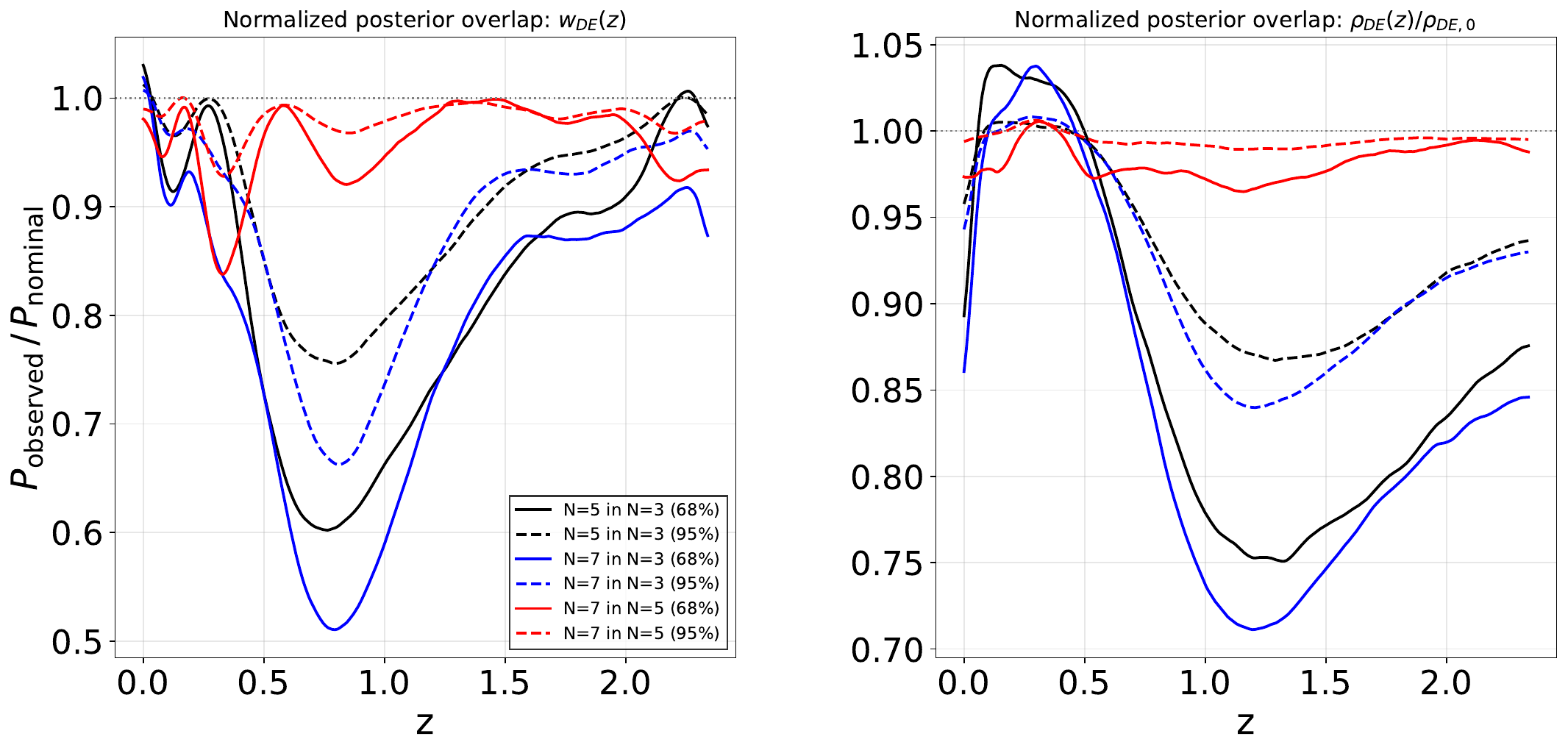}
\caption{Posterior overlap for the $w_{DE}\left( z\right) $ and $\protect%
\rho _{DE}\left( z\right) $ for reconstructed functions from different
orders of the Chebyshev polynomial. The comparison are for $N=3,$ $5~$and $7$%
.}
\label{fig5}
\end{figure}

In Fig. \ref{fig4} we present the trained cosmological solution for the same
set of initial conditions and different degrees of the Chebyshev polynomials
for the combined data CC+BAO+PP. Specifically we trained the network for $%
N=3,$\ $5\ $and $7$. Recall that for $N=1$, the Chebyshev polynomial is the
linear function. In Fig. \ref{fig5} we present the overlap of the posterior
bands for the $N=5$\ to $N=3\,$, for the $N=7$\ to $N=3$\ and $N=7$\ with
respect to $N=5$, for the $68\%$\ and $95\%$\ credible intervals. In Table %
\ref{cheb} we summarize the mean values for the normalized posterior overlap
as presented in Fig. \ref{fig5}. We observe that while there is a
significant difference from $N=3$\ to $N=5$, the difference in the
reconstruction from $N=5$\ to $N=7$\ is not statistically significant. ~We
see that the Cosmo-PINN is stable and the results statistically equivalent
for the choice of degree for the Chebyshev expansion for $N\geq 4$.

Furthermore, in order to understand how the different choices of $N$\ affect
the statistics, we compute the $AIC$\ of the likelihoods, from which we
obtain the Akaike Information Criterion~$\Delta \left( AIC\right)
_{5-3}\simeq 4$ and $\Delta \left( AIC\right) _{7-5}\simeq 4$. The goodness
of fit has already reached its limit at $N=3$, so that the additional
coefficients of the Chebyshev expansion contribute only to the complexity.
Thus, the Akaike penalty $2k$\ presumes that every additional parameter is
an effective degree of freedom, which is not the case for coefficients that
are consistent with zero in a near-orthogonal basis. A similar behaviour for
the AIC reported before in \cite{mpp1}. However, in \cite{mpp1} it has been
considered a monomial basis, $(1-a)^{l}$, for the expansion of the dark
energy equation of state parameter.\ The monomial basis is not orthogonal,
which leads to a degeneracy between the coefficients. For an orthogonal
basis, as for the Chebyshev expansion, the degeneracy is removed, so that
increasing $N$\ introduces independent information rather than
redistributing it among the existing coefficients. This is observed in Table %
\ref{cheb}, and it is the reason why a single sufficiently large $N$\ is
sufficient for this study.

\begin{table}[tbp] \centering%
\caption{Mean normalized posterior overlap of the reconstructed $w_{DE}(z)$
and $\rho_{DE}(z)$ for different Chebyshev expansions.}%
\begin{tabular}{ccc}
\hline\hline
\textbf{Comparison} & \textbf{CI} & $\left\vert \frac{\Delta P}{P}%
\right\vert $ \\ \hline
\multicolumn{3}{c}{$\mathbf{w}_{DE}\left( z\right) $} \\ \hline
$N=5$ in $N=3$ & $68\%$ & $17.7\%$ \\ 
$N=5$ in $N=3$ & $95\%$ & $9.9\%$ \\ 
$N=7$ in $N=3$ & $68\%$ & $21.6\%$ \\ 
$N=7$ in $N=3$ & $95\%$ & $12.3\%$ \\ 
$N=7$ in $N=5$ & $68\%$ & $4.2\%$ \\ 
$N=7$ in $N=5$ & $95\%$ & $1.9\%$ \\ \hline
\multicolumn{3}{c}{$\mathbf{\rho }_{DE}\left( z\right) \mathbf{/\rho }%
_{DE,0} $} \\ \hline
$N=5$ in $N=3$ & $68\%$ & $13.6\%$ \\ 
$N=5$ in $N=3$ & $95\%$ & $7.3\%$ \\ 
$N=7$ in $N=3$ & $68\%$ & $15.8\%$ \\ 
$N=7$ in $N=3$ & $95\%$ & $8.2\%$ \\ 
$N=7$ in $N=5$ & $68\%$ & $1.8\%$ \\ 
$N=7$ in $N=5$ & $95\%$ & $0.5\%$ \\ \hline\hline
\end{tabular}%
\label{cheb}%
\end{table}%

\begin{figure}[tbph]
\centering\includegraphics[width=1\textwidth]{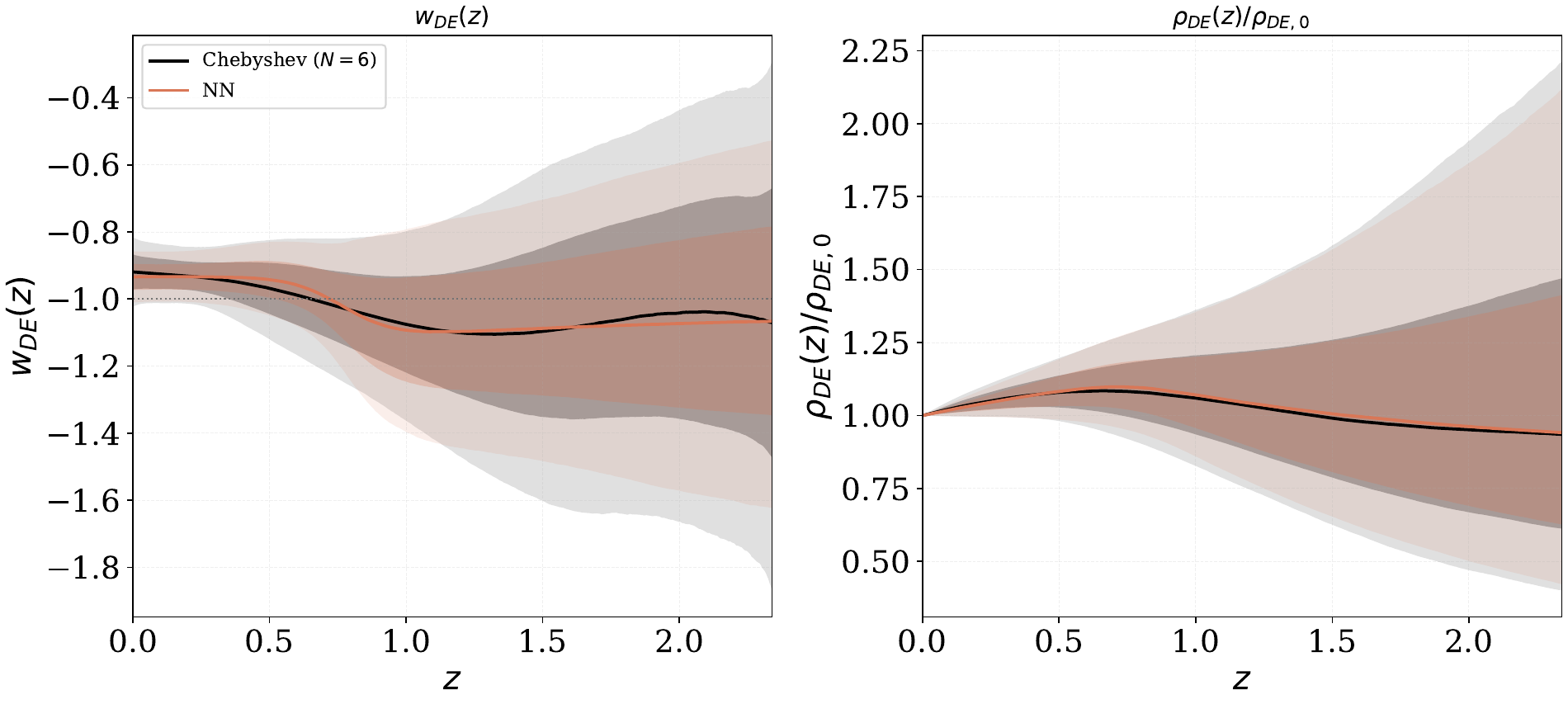}
\caption{Reconstructed dark energy equation of state parameter $w_{DE}\left(
z\right) $ and $\protect\rho _{DE}\left( z\right) $ for the two different
representations of the $w_{DE}\left( z\right) ~$for the combined data
CC+BAO+PP. Specifically, for the Chebyshev representation for $N=6$, and the
neural network representation in which $w_{DE}\left( z\right) $ is given by
a two hidden layer network with 32 neurons each.}
\label{fign1}
\end{figure}

\begin{figure}[tbph]
\centering\includegraphics[width=1\textwidth]{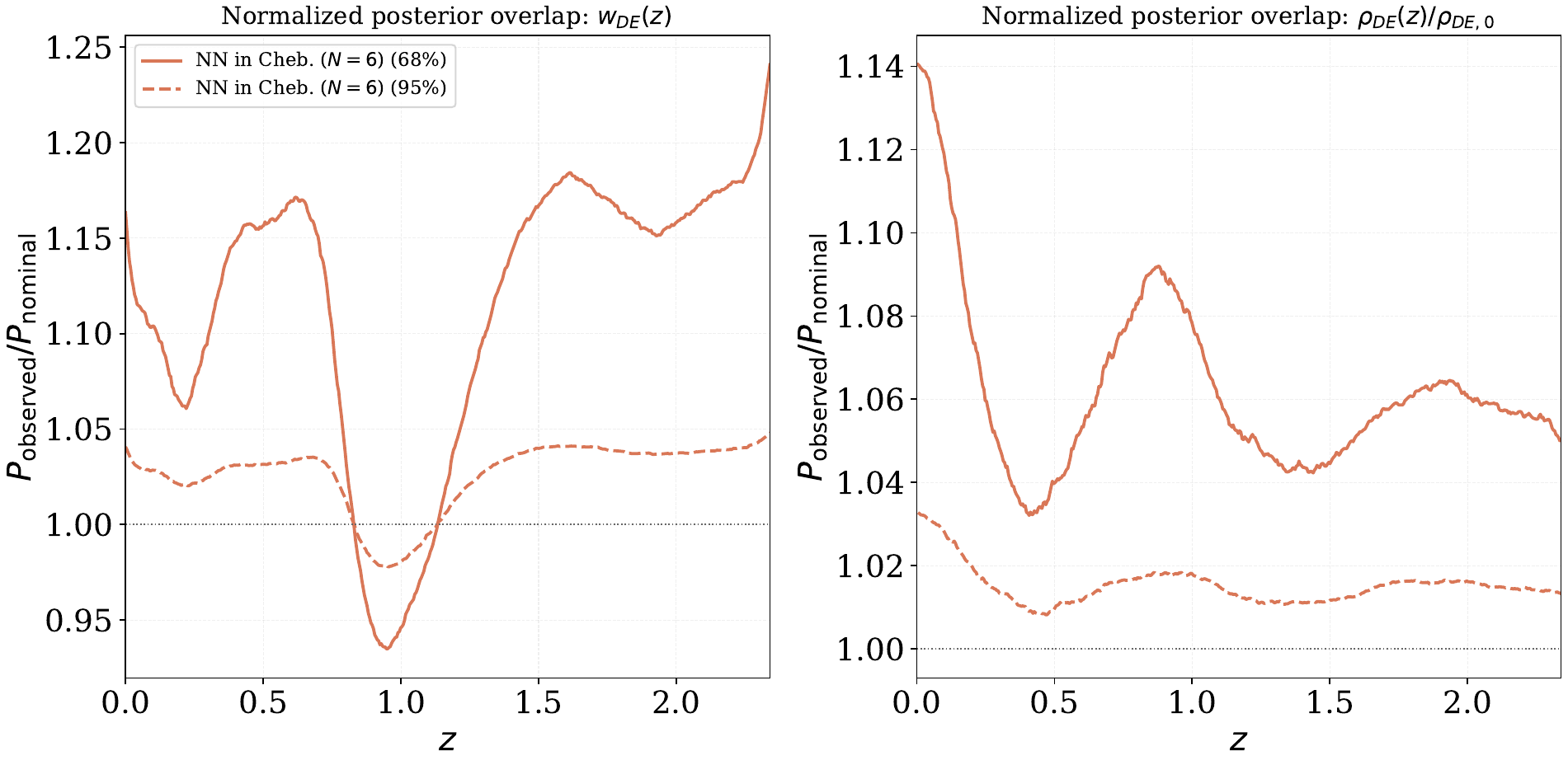}
\caption{Posterior overlap for the $w_{DE}\left( z\right) $ and $\protect%
\rho _{DE}\left( z\right) $ for reconstructed functions the Chebyshev
representation for $N=6$, and the neural network representation in which $%
w_{DE}\left( z\right) $ is given by a two hidden layer network with 32
neurons each. }
\label{fign2}
\end{figure}

The comparison of the reconstructed $w_{DE}\left( z\right) $\ for the neural
network representation, with the Chebyshev representation for $N=6$, is
presented in Fig. \ref{fign1}, while the normalized posterior overlap is
presented in Fig. \ref{fign2}. The mean values of the posterior overlap are
summarized in Table \ref{chebnn}

\begin{table}[tbp] \centering%
\caption{Mean normalized posterior overlap of the reconstructed $w_{DE}(z)$
and $\rho_{DE}(z)$ for the Chebyshev and the Neural Network representations for the $w_{DE}(z)$.}%
\begin{tabular}{ccc}
\hline\hline
\textbf{Comparison} & \textbf{CI} & $\left\vert \frac{\Delta P}{P}%
\right\vert $ \\ \hline
\multicolumn{3}{c}{$\mathbf{w}_{DE}\left( z\right) $} \\ \hline
Network in Chebyshev $\left( N=6\right) $ & $68\%$ & $12.5\%$ \\ 
Network in Chebyshev $\left( N=6\right) $ & $95\%$ & $3.0\%$ \\ \hline
\multicolumn{3}{c}{$\mathbf{\rho }_{DE}\left( z\right) \mathbf{/\rho }%
_{DE,0} $} \\ \hline
Network in Chebyshev $\left( N=6\right) $ & $68\%$ & $6.2\%$ \\ 
Network in Chebyshev $\left( N=6\right) $ & $95\%$ & $1.5\%$ \\ \hline\hline
\end{tabular}%
\label{chebnn}%
\end{table}%

\begin{figure}[tbph]
\centering\includegraphics[width=1\textwidth]{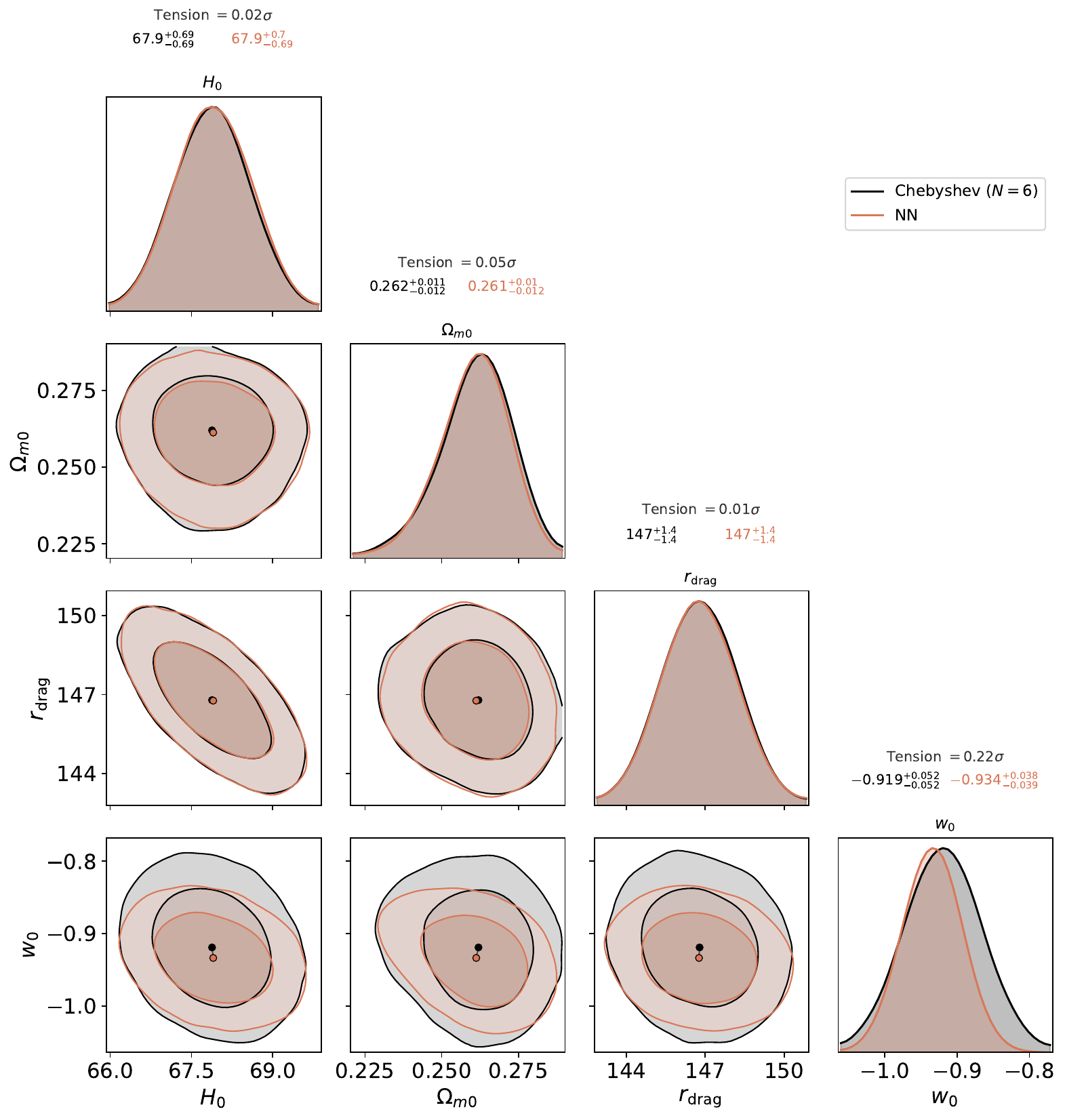}
\caption{Confidence contours for the trained cosmological parameter $H_{0}$,~%
$\Omega _{m0}$, $r_{drag}$ and the present value of the dark energy equation
of state parameter $w_{DE}\left( z\rightarrow 0\right) =w_{0}$, as obtained
from the reconstruction by representing the $w_{DE}\left( z\right) $ by a
neural network, or a Chebyshev expansion of sixth-order. The median values
and the tension of the two representations are reported. }
\label{fgn3}
\end{figure}

Moreover, Fig. \ref{fgn3} shows the confidence contours for the trained
cosmological parameters $H_{0},~\Omega _{m0},$\ $r_{drag}$\ and the obtained
present value for the $w_{DE}\left( z\rightarrow 0\right) =w_{0}$\ as given
by the two different representations for the $w_{DE}\left( z\right) $. We
observe that within the cosmological parameters the tension~$\left( T\right) 
$\ is $T\leq 0.05\sigma $, while for the present value of the $w_{DE}$, the
tension is $T=0.22\sigma $. The reconstruction of the background is
therefore independent of the parametrization for these datasets. The wider
errors of the $w_{DE}\left( z\right) $\ at the present are due to the
contribution of all the coefficients of the Chebyshev expansion at the
boundary.

\section{Reconstruction of Teleparallel $f\left( T\right) $ Gravity}

\label{append}

To demonstrate the application of Cosmo-PINN to alternative theories of
gravity, we consider the reconstruction of teleparallel $f\left( T\right) $%
-gravity \cite{ft,ft2}. \ Within the spatially flat FLRW geometry (\ref%
{cc.01}) and for the vierbein field 
\begin{equation}
h_{\mu }^{i}(t)=diag(1,a(t),a(t),a(t)),
\end{equation}%
the torsion scalar reads%
\begin{equation}
T=-6\left( \frac{\dot{a}}{a}\right) ^{2}=-6H^{2},  \label{ft.00}
\end{equation}%
and Friedmann's field equations are modified as \cite{ft3}%
\begin{eqnarray}
12H^{2}f_{T}(T)+f(T) &=&2\left( \rho _{m}+\rho _{b}+\rho _{r}\right) ,
\label{ft.01} \\
48H^{2}\dot{H}f_{TT}(T)-4(\dot{H}+3H^{2})f_{T}(T)-f(T) &=&2p_{r},
\end{eqnarray}%
where the matter component satisfies the continuous equations (\ref{c.10}), (%
\ref{c.11}) and (\ref{c.12}).

Equation (\ref{ft.01})\ can be expressed in the equivalent form%
\begin{equation}
3H^{2}=\rho _{m}+\rho _{b}+\rho _{r}+\rho _{f}  \label{ft.02}
\end{equation}%
in which $\rho _{f}=Tf_{T}-\frac{1}{2}\left( f+T\right) ~$or equivalently 
\cite{anbarrow}%
\begin{equation}
\rho _{f}=TF_{,T}-\frac{1}{2}F.  \label{ft.04}
\end{equation}%
where we have replaced $f\left( T\right) =T+F\left( T\right) $. $\rho _{f}~$%
is the dark energy component related to the nonlinear function $f\left(
T\right) $. We remark that for $f\left( T\right) =T-2\Lambda $,~that is $%
F=-2\Lambda .$\ the limit of General Relativity with a cosmological constant
is recovered, while a component $\sqrt{-T}$\ in the function $f\left(
T\right) \rightarrow f\left( T\right) +\beta \sqrt{-T}$, leads to the same
field equations, because the $\sqrt{-T}$\ component is total derivative for
the FLRW background.

We define the dimensionless variable~$\mathcal{T}=-\frac{T}{6H_{0}^{2}}%
=\left( \frac{H}{H_{0}}\right) ^{2},$\ then we can write the energy density
in the following form%
\begin{equation}
\bar{\rho}_{f}=\Omega _{T,0}\left( \bar{F}\left( \mathcal{T}\right) -2%
\mathcal{T}\bar{F}_{\mathcal{T}}\left( \mathcal{T}\right) \right) ,
\label{ft.05}
\end{equation}%
where $F\left( \mathcal{T}\right) =\Omega _{T,0}F\left( \mathcal{T}\right) $%
. Moreover we consider $F\left( 1\right) =1$\ and $F_{\mathcal{T}}\left(
1\right) =0$, such that to avoid the appearance of the component $\sqrt{-T}$%
\ in the reconstructed function and the effective gravitational constant to
be consistent with that of General Relativity,~that is $f_{,T}\left(
1\right) =1$.

We modify the Cosmo-PINN framework presented in Section \ref{sec3} in order
to train the network for the $f\left( T\right) $-theory.

The loss term of the physics is defined now by the residual 
\begin{equation}
\mathcal{R}_{T}:\bar{\rho}_{f}-\Omega _{T,0}\left( \mathcal{F}\left( 
\mathcal{T}\right) -2\mathcal{TF}_{\mathcal{T}}\left( \mathcal{T}\right)
\right) ,~\mathcal{T}=E^{2}\left( x\right) ,  \label{ft.06}
\end{equation}%
where function $F\left( \mathcal{T}\right) $\ is described by a second
network, or it can be expressed in term of Chebyshev polynomial, similar to
the $w_{DE}\left( z\right) $. The dark energy equation of state parameter is
defined as \cite{anbarrow} 
\begin{equation}
w_{f}=-\frac{\mathcal{T}-\mathcal{TF}_{\mathcal{T}}+2\mathcal{TF}_{\mathcal{%
TT}}}{\left( 1+\mathcal{F}_{\mathcal{T}}+2\mathcal{TF}_{\mathcal{T}}\right)
\left( \mathcal{F}-2\mathcal{TF}_{\mathcal{T}}\right) }.
\end{equation}%
Furthermore, the regularity conditions are modified to ensure the smoothness
of the reconstructed function$~F\left( \mathcal{T}\right) $, while the
condition $f_{,T}>0$, which is necessary for the viability of the theory is
introduced \cite{ft3}.

\begin{figure}[h]
\centering\includegraphics[width=1\textwidth]{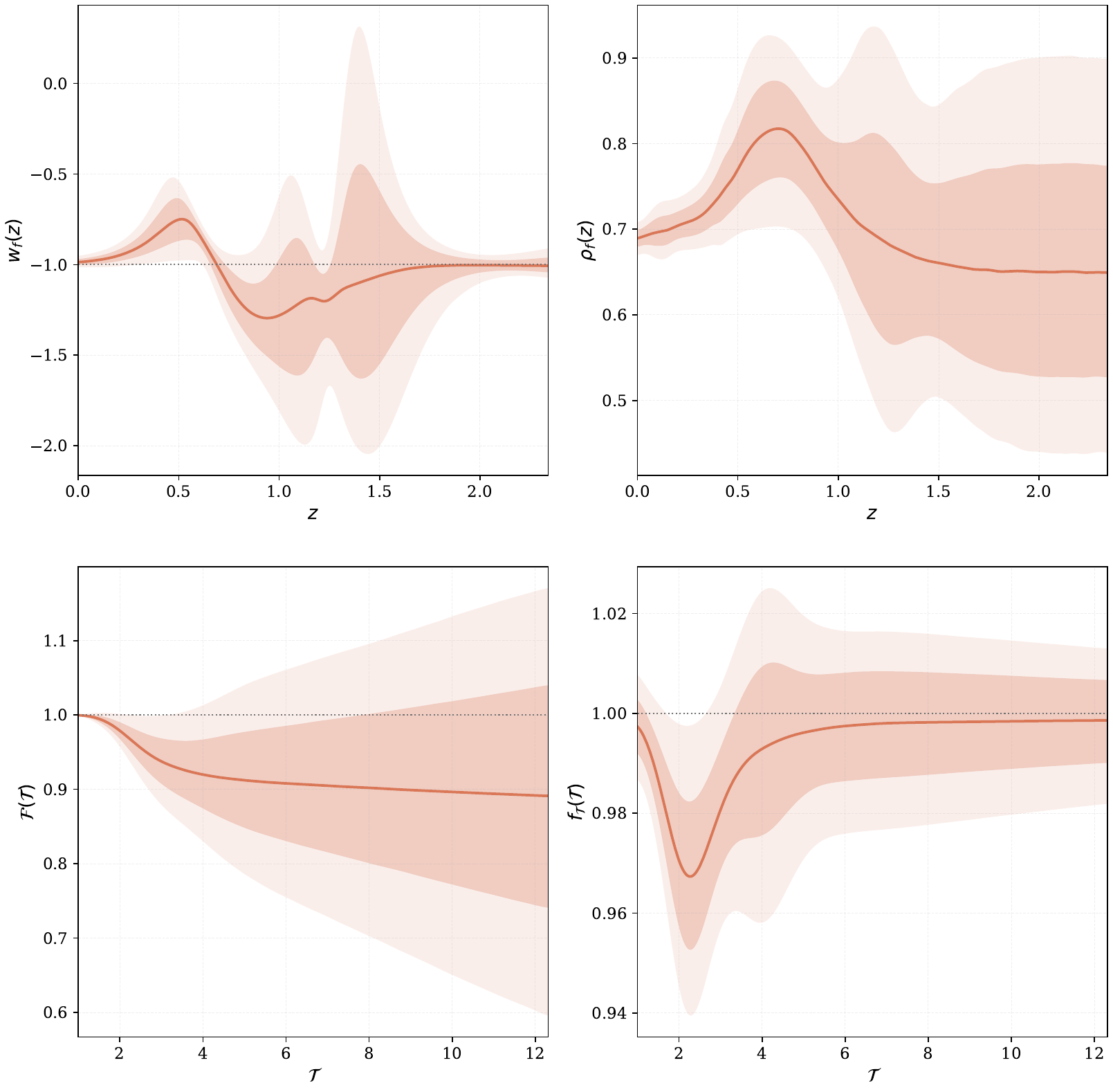}
\caption{Cosmo-PINN\ output for the dark energy equation of state parameter $%
w_{f}\left( z\right) $, dark energy density $\protect\rho _{f}\left(
z\right) $, teleparallel $\mathcal{F}\left( \mathcal{T}\right) $ and the
first derivative $f_{\mathcal{T}}\left( \mathcal{T}\right) $, for the CC+BAO
combined data, where the $\mathcal{F}\left( \mathcal{T}\right) $ is
represented by a $2\times 32$ neural network.}
\label{figft}
\end{figure}

We perform the reconstruction of the teleparallel $F\left( \mathcal{T}%
\right) $\ for the combined data CC+BAO, and where the function $F\left( 
\mathcal{T}\right) $\ is represented as a neural network of two layers with
32 neurons each layer. The training take place with the requirement $\rho
_{T}>0$. The results of the Cosmo-PINN are presented in Fig. \ref{figft}.
The reconstruction indicate a small deviation from the $\Lambda $CDM, where
for large redshifts, the $\Lambda $CDM limit is recovered. The evolution of
the obtained effective dark energy is different from that found before, that
is, because of the physical constraints introduced to the theory. \\

\textbf{Data Availability Statement:} The source code of Cosmo-PINN is publicly available at https://github.com/anpaliat/Cosmo-PINN. \\ 

\begin{acknowledgments}
The author thanks the anonymous referee for the valuable comments and
suggestions that helped improve this study. This work was partially
supported by FONDECYT Grant 1240514. The training of the Cosmo-PINN was
carried out using the computing cluster of the Universidad Catolica de
Temuco and GPU resources provided by Vast.ai.
\end{acknowledgments}

\end{document}